\documentclass[12pt]{article}
\usepackage{epsfig}
\newcommand{\be}{\begin{equation}}
\newcommand{\ee}{\end{equation}}
\newcommand{\bea}{\begin{eqnarray}}
\newcommand{\eea}{\end{eqnarray}}
\newcommand{\ta}{\tilde\alpha}
\newcommand{\tb}{\tilde\beta}

\newcommand{\bm}{\bar\mu}
\newcommand{\bn}{\bar\nu}
\newcommand{\da}{\dot\alpha}
\newcommand{\db}{\dot\beta}
\newcommand{\la}{\lambda}
\newcommand{\ga}{\gamma}
\newcommand{\al}{\alpha}
\newcommand{\bet}{\beta}
\newcommand{\e}{\eta}
\newcommand{\GSM}{$SU(3)\times SU(2)_L \times U(1)_Y\quad $}
\newcommand{\PS}{$ SU(4)\times SU(2)_L\times SU(2)_R\quad  $}

\newcommand{\et}{\tilde \eta}
\newcommand{\mht}{{\widetilde M}_H}

\newcommand{\at}{\tilde a}
\newcommand{\pt}{\tilde p}
\newcommand{\omt}{\tilde \omega}
\newcommand{\om}{\omega}
\newcommand{\sgt}{\tilde \sigma}
\newcommand{\sgbt}{\tilde {\overline{ \sigma}}}
\newcommand{\Sig}{\bf\Sigma}
\newcommand{\sss}{\sigma}
\newcommand{\ssb}{{\overline{ \sigma}}}
\newcommand{\sq}{\sqrt{2}}
\newcommand{\sqs}{\sqrt{6}}
\newcommand{\sqt}{\sqrt{3}}
\newcommand{\sqf}{\sqrt{5}}
\newcommand{\sqtt}{\sqrt{3\over 2}}
\newcommand{\os}{\overline\Sigma}
\newcommand{\s}{\Sigma}
\newcommand{\Sigb}{{\overline\Sigma}}
\newcommand{\oot}{\overline {126}}
\newcommand{\bh}{\bar h}
\newcommand{\bt}{\bar t}
\newcommand{\ovt}{\overline{10}}
\newcommand{\ovl}{\overline}
\begin{document}

 \vfil
 \vspace{3.5 cm} \Large{
\title{\bf  {SO(10) MSGUT : Spectra,
 Couplings and Threshold Effects}}
 \author{ Charanjit S. Aulakh  and  Aarti Girdhar}}
\date{}
\maketitle



 \normalsize\baselineskip=15pt

 {\centerline  {\it
Dept. of Physics, Panjab University}} {\centerline{ \it
{Chandigarh, India 160014}}}

 {\centerline {\rm E-mail: aulakh@pu.ac.in }}
\vspace{1.5 cm}

\large {\centerline{\bf {ABSTRACT }}}
\normalsize\baselineskip=15pt

\vspace{1. cm}

 { We compute  the complete  gauge and
chiral superheavy mass spectrum and couplings of the Minimal Susy
GUT  (based on the  $ \bf {210- \oot- 126-10 }$ irreps as the
Higgs system) by decomposing    SO(10) labels   in terms of Pati
Salam subgroup labels. The spectra are sensitive functions of the
single complex parameter that controls MSGUT symmetry breaking. We
scan
  for the dependence of the threshold corrections to the
Weinberg angle and Unification scale as functions of this
parameter.
 We find that for
generic values of the GUT scale parameters the modifications are
within 10\% of the one loop values and can be much smaller for
significant regions of the parameter space. This shows that
contrary to longstanding conjectures,  high precision calculations
are not futile but rather   necessary and  feasible   in the
MSGUT.
 The couplings of the matter supermultiplets
 are made explicit and used to identify the
 channels for exotic ($\Delta B\neq 0$) processes
 and to write down the associated bare $d=5$ operators
 (some of both are novel).
The mass formulae for all matter fermions are derived.
 This sets the stage for a comprehensive
  RG based phenomenological analysis of the MSGUT.}

\normalsize\baselineskip=15pt

\section{ Introduction}

The  Supersymmetric SO(10) GUT  based on the ${\bf 126}, {\bf
{\overline{126}}},
 {\bf 210}$   Higgs multiplets
\cite{aulmoh,ckn,lee,abmsv}  has, of late, enjoyed a much delayed
bloom of interest motivated by its economy and predictivity.
Besides the traditional virtues of SO(10) this is the minimal
renormalizable model   which has shown itself capable of matching
the observed fermion spectra, including the prima facie GUT
repellent feature of maximal mixing in the neutrino sector
\cite{babmoh,bsv,moh}.
 Beyond  the traditional scenario of perturbative unification of
couplings due to the RG flow between  $M_S$ and $M_X$  it also
offers strong indications that the gauge coupling becomes strong
above the GUT scale.
  We have argued \cite{trmin,tas}
that this necessarily  leads to dynamical symmetry breaking of the
GUT symmetry at a scale $\Lambda_U$ (just above the perturbative
unification scale $M_U\sim 10^{16} GeV $). Utilizing the
quasi-exact supersymmetry at the GUT scale we made
plausible\cite{tas}
  a scenario in which $\Lambda_U$ is {\it{calculably}}
determined by only the low energy data and structural features of
the theory (such as
 the gauge symmetry group, supersymmetry and the very restricted
Higgs multiplets available to generate fermion masses
-particularly neutrino masses- in a renormalizable  theory). This
scenario offers interesting
 possibilities of a novel  picture of elementarity and dual unification characterized by
a new fundamental length scale $\sim \Lambda_U^{-1}$
 characterizing
 the ``hearts of   quarks.''\cite{trmin,tas}.

The MSGUT is thus the focus of multi-faceted interest and a
 detailed phenomenological analysis of the theory in terms of
the structure dictated by its GUT scale minimality is thus called
for. However such an analysis has been   delayed by the
computational difficulty of obtaining the GUT scale spectra and
couplings and the
 effective Lagrangian describing the normal and
exotic features (baryon and lepton violation etc) of the GUT
derived MSSM (i.e extended by the leading ($d=5$)
 exotic operators of the theory).
The spectra and couplings are necessary   to analyse threshold
corrections to the gauge couplings near the GUT scale and are also
a crucial input into deriving the lagrangian for exotic processes
and parameters   mediated by GUT scale massive fermions. In
\cite{alaps} we presented  techniques for computing the
decomposition of SO(10) invariants in terms of the unitary labels
of  its maximal (Pati-Salam) sub group \PS. Once this
decomposition is performed the computation of the complete
spectrum and couplings is quite easy  and the long standing
vagueness  regarding the ``Clebsches'' that arise can finally be
banished. This allowed us to present, by way of illustration of
the power of our method, the two most important mass matrices ($4
\times 4$ and $5\times 5$ respectively) affecting  Electro weak
symmetry breaking, fermion masses and nucleon decay :
  namely those for the MSSM type Higgs $SU(2)_L $ doublets
and baryon number violation mediating $SU(3)_c$ colour triplets
 that mix with the the doublets and
triplets of the fermion mass (FM) Higgs (${\bf{10, \oot}}$).
 Moreover since our methods allow computation of the actual
couplings of Higgs to spinors  we could also obtain the $d=5 $
operators for Baryon violation generated via exchange of triplet
Higgsinos contained not only in the traditional $(6,1,1)$
submultiplets (of the ${\bf{10}}$  or those in the ${\bf{\oot}}$
\cite{babpatwil}) which had been noticed to provide  a connection
between neutrino masses and proton decay, but also in other
channels arising from the exchange of colour triplets contained in
$(10,1,3)_{\oot}$ submultiplets involved in neutrino mass
generation\cite{alaps}.
     A more complete calculation of these spectra and effective lagrangians
 and an initial estimate of their effects
 is the subject of this paper.

While the calculations presented were in their final stages
  we were collaborating and cross checking   with another
parallel  calculation\cite{bmsv} of mass spectra using a different
 \cite{heme}
method which has since been published .
 Moreover another group \cite{fm,fmv2}
has also recently published a calculation
 (using the same methods as \cite{bmsv})
 of spectra and baryon decay effective potentials
recently. As far as computation of chiral superfield spectra are
concerned our results coincide (upto normalization and phase
conventions) with those of \cite{bmsv}. However both  our results
diverge\cite{alaps,bmsv}  in certain details from the chiral
spectra given in \cite{fm}. Moreover as already noted by us in (an
update to) \cite{alaps} we
 also {{disagreed}} with the results
of \cite{fm}  regarding the Higgsino channels available for baryon
decay in this model.  We found \cite{alaps} that \cite{fm}
obtained couplings between the ${\bf\oot}$ multiplet and matter in
the spinorial $\bf{16}$
 representation which were in contradiction with ours\cite{alaps}
 not only as regards the numerical coeffcients but also in the
heavy Higgs channels to which matter fields couple in a baryon
number violating way. In the revised version of \cite{fm} i.e
\cite{fmv2} this defect has apparently been corrected at least
modulo disagreement on values of clebsches. We shall try to settle
these  questions by tracing the reasons for the continuing
discrepancy in explicit detail and  confirm our previous
assertions.
 We have also
analyzed the gauge Dirac multiplet structure arising from the
super-Higgs effect and the masses and vevs responsible for the
Type I and Type II mechanisms \cite{seesaw,mohsen} of neutrino
mass generation.

We emphasize that our method allows computation,
 not only of
spectra but also of the  couplings  of all  the multiplets  in the
theory (whether they are renormalizable or  heavy-exchange induced
effective couplings) without any ambiguity. Moreover our results
are obtained by  an analytic tensorial reprocessing of labels of
fields in the Lagrangian.
 This approach might thus
 find  preferment with field theorists
 in comparison with the more restricted capabilities
   of the approach of
\cite{heme}, which, so far, has not proved capable
 of generating all the Clebsches of the SO(10)
theory and which  relies on an explict multiplet
 representative and computer based approach
which is  tedious to connect to the    unitary group tensor
methods so familiar to particle theorists.

It has long been held by some that $SO(10) $ GUTs specially Susy
$SO(10)$ GUTs, are essentially self-contradictory \cite{dixitsher}
due to the apparently enormous threshold effects that might arise
due to the large number of superheavy Higgs residuals in these
theories. Thus the authors of \cite{dixitsher} speak of the the
``futility of high-precision calculations in SO(10)". However
these assertions have never been tested against any actual
computations of mass spectra of a Susy SO(10) GUT and are only
worst case estimates assuming that no cancellations
 occur. However
we expect that cancellations {\it{will}} generically occur since
the lepto-quark mass has no reason to lie at the edge of the mass
spectrum nor are the coefficients all of the same sign.
 With the
computed spectra now available, the threshold effects on
observable quantities such as $Sin^2\theta_w, M_X$ etc become
computable  in terms of the relatively small number \cite{abmsv}
of GUT scale parameters of the MSGUT. In fact \cite{abmsv} a
{\it{single}} complex parameter controls the solutions of the
cubic equation in terms of which all the superheavy vevs are
defined. We have performed a preliminary scan of the parameter
space of the MSGUT to see what is the typical size of these
corrections. We find the striking result that such threshold
corrections are {\it{generically}} $\leq 10\%$ of the 1-loop
results\cite{geoquinwein,marsenj,jones,amaldi} that underpin the
 the GUT scenario's viability. Moreover, for significant and
 possibly interestingly restricted regions of
 parameter space these corrections  can be
 {\it{much smaller}} i.e as small as .5\%
 of the one loop results.  Thus far from indicating
  futility our results indicate that a thoroughgoing investigation
  of the  compatibility of low energy precision
  data with the   threshold corrections may
  significantly constrain the parameter
  space of the MSGUT. In any case we show that inclusion
  of threshold corrections is necessary and not futile.
  Such an analysis is now in progress \cite{abmsv2}.

               In Section  2. we present a brief review of the principal features of
 the minimal Susy SO(10) theory \cite{aulmoh,ckn,lee,abmsv} and
compute the gauge supermultiplet masses.
 In Section 3. we provide
the PS reduction of the SO(10) Higgs superpotential. From this we
computed the Chiral fermion mass terms and thus the supermultiplet
spectra which we discuss here and list in an Appendix .
  In Section  4. we  compute the threshold corrections to the
  1-loop values of the unification scale $M_X$ and $sin^2\theta_W(M_S)$.
   In Section 5. we
present the couplings of the matter fields to FM Higgs fields in
the superpotential as well as their couplings to  the gauginos of
the SO(10) model. This permits us to identify the possible
channels for baryon violation in the low energy theory via
exchange of Higgsinos or gauginos and compute the relevant
effective lagrangians. Using the associated mass matrices we write
down the $d=5$ effective lagrangians for baryon and lepton number
violation
 which arise via exchange of superheavy fermions.
In Section 6. we discuss the mass formulae for the matter fermions
in  this model.
  The  majorana mass terms  of the left and right handed neutrinos
and  the $SU(2)_L$ triplet  micro -vev
 responsible the Type II mechanism for
neutrino mass is calculated along with the charged fermion mass
matrices.    In a final section we discuss our conclusions and
results and  plans for further investigations using the results
derived here.

 \section{The Minimal Susy GUT }
 In accordance with our basic rationale we shall
deal with a renormalizable
 globally supersymmetric $SO(10)$ GUT
 whose chiral supermultiplets  consist of ``adjoint multiplet type''
(or AM) totally antisymmetric tensors : $ {\bf{210}}(\Phi_{ijkl}),
{\bf{\overline{126}}}({\bf{\Sigb}}_{ijklm}),$
 ${\bf{126}} ({\bf\Sigma}_{ijklm}) \hfil\break (i,j=1...10)$
which serve to break the GUT symmetry
 to the SM, together with Fermion mass (FM) Higgs
{\bf{10}}-plet(${\bf{H}}_i$).
  The  ${\bf{\overline{126}}}$ plays a dual or AM-FM
role since besides enabling Susy preserving GUT symmetry breaking,
it also enables the generation of realistic charged fermion masses
  and   neutrino masses and mixings (via the Type I
and/or Type II mechanisms) ;
 three spinorial {\bf{16}}-plets ${\bf{\Psi}_A}(A=1,2,3)$  contain the
matter supermultiplets together with the three conjugate neutrinos
(${\bar\nu_L^A}$).
  The ${\bf{\overline{126}}}$
  $({\bf{\overline{\Sigma}}} ),
 {\bf{126}}({\bf{\Sigma}}) $
pair is required to be present together to preserve Susy while
 breaking  $U(1)_{R}\times U(1)_{B-L}\rightarrow U(1)_{Y}$
and is capable{\cite{babmoh,bsv,moh}} of generating realistic
neutrino masses and mixings via the type I or type II seesaw
mechanisms {\cite{seesaw,mohsen}}.
 The complete superpotential in this theory
 is the sum of

 \bea W_{AM}&=&{1\over{2}}M_{H}H^{2}_{i} + {m \over
{ 4! }} \Phi_{ijkl}\Phi_{ijkl}+{\lambda \over
{{4!}}}\Phi_{ijkl}\Phi_{klmn} \Phi_{mnij}+{M \over { 5!
}}\Sigma_{ijklm}\overline\Sigma_{ijklm}\nonumber\\ &+&{\eta \over
4!}\Phi_{ijkl}\Sigma_{ijmno}\overline\Sigma_{klmno}+
 {1\over{4!}}H_{i}\Phi_{jklm}(\gamma\Sigma_{ijklm}+
\overline{\gamma}\overline{\Sigma}_{ijklm})\label{WAM} \eea

and  \be W_{FM}=
h_{AB}'\psi^{T}_{A}C^{(5)}_{2}\gamma_{i}\psi_{B}H_{i}+ {1\over
5!}f_{AB}' \psi^{T}_{A}C^{(5)}_{2}{\gamma_{i_{1}}}...
{\gamma_{i_5}}\psi_{B}\overline{\Sigma}_{i_1...i_5}\label{FMpot}
 \label{WFM}\ee
Our notations and conventions for spinors can be found in
\cite{alaps}. The Yukawa couplings $h_{AB}',f_{AB}'$ are complex
symmetric $3\times 3 $ matrices and one of them, say $f'$ can thus
be diagonalized (by an orthogonal transform $UfU^T$ using a
Unitary matrix $U$ which leaves the matter kinetic terms
invariant) to a real positive diagonal form $f'_{AB}=F'_A
\delta_{AB}$, thus leaving 15 residual real parameters in
$W_{FM}$.
 In addition  the 7 complex parameters in $W_{AM}$ can be reduced
to 10 real ones by absorbing 4 phases by Higgs field
redefinitions. Then together with the gauge coupling one has in
all exactly 26 non-soft parameters \cite{abmsv}. Coincidentally,
MSSM also has 26 non-soft couplings consisting of 9 quark and
charged lepton masses, 3 majorana neutrino masses 3 quark (CKM)
and 3 lepton (PMNS) mixing angles and 1 quark but 3 lepton  CP
phases together with 3 gauge couplings and a $\mu$ parameter. Thus
we see that the 15 parameters of $W_{FM}$ must be essentially
responsible for the 22 parameters describing fermion masses and
mixings in the MSSM.

The kinetic terms  are given by covariantizing in the standard way
the global SO(10) invariant D-terms

\be
 [{1\over {2.5!}}(\Sigma_{ijklm}^*  \Sigma_{ijklm}
  +  {\overline\Sigma}_{ijklm}^*   {\overline\Sigma}_{ijklm})  +
{1\over{4!}}\Phi_{ijkl}^*  \Phi_{ijkl} + H_i^*  H_{i}  + \Psi_A^*
\Psi_A]_D \ee

Note that the extra  factor of (1/2) achieves canonical
normalization for the 126 independent component fields  of the
self dual ( anti self-dual )
 ${\bf{126}}({\bf{\overline{126}}})$ representations  which
would be otherwise be overcounted.
 We thus, unfortunately,  differ from the  normalization used in the parallel computations of
\cite{bmsv} with which our results are nevertheless in agreement
after appropriate rescalings and rephasings of parameters and
fields.  We emphasize that all our redefinitions of labels are
unitary and thus maintain unit norm relative to the above kinetic
terms.\footnote{The relations between the quantities of
\cite{bmsv}(denoted by primes) and
 ours are $\Sigma'=\Sigma/{\sqrt 2},
 {\overline{\Sigma}}'={\overline{\Sigma}}
  /{\sqrt 2}$ (also for vevs)
 $ \gamma'={\sqrt 2} \gamma,  {\bar\gamma}'=
 {\sqrt 2} {\bar{\gamma}}, \eta'=2\eta,M'=2M $.}

The economy of the above superpotential
eqns.(\ref{WAM},\ref{FMpot})   is remarkable\cite{abmsv}. It's few
couplings together with the functional flexibility of the chosen
Higgs multiplet set    and   its ability (in common with other
renormalizable models using just $10-\oot$ FM Higgs) to fit the
all the fermion mass data\cite{bsv,moh}, justify its claim to
being {\it{the}} MSGUT.
 The ``small" number of non-soft parameters (26 as in the MSSM) implies that
after fitting \cite{babmoh,bsv,moh} the known
 quark, charged lepton masses and quark mixing angles together
with the neutrino mass splittings very little play is left in
 the model and it becomes predictive and thus falsifiable.
The nearest related model(NMSGUT?)(in some ways more logically
complete since all the FM channels allowed by renormalizability
would then be utilized)  might be considered to be  the one
obtained by adding a 120-plet SO(10) FM Higgs. Alternatively one
may consider SU(5) supplemented with right handed neutrinos or
non-renormalizable terms \cite{abmsv}. Both models
 are are far less economical and  not so predictive. Therefore,
as advocated  in  detail in \cite{abmsv},
 the first priority should be to pin down
the predictions of  this model. We began the development of a
detailed framework for handling the group theoretic complexity of
susy SO(10)
 models generally in
\cite{alaps} and this paper presents the results of
 calculations using the techniques developed there for
computing couplings and spectra for MSSM fields from the MSGUT
tree action by decomposing the fields according to the \PS  or
Pati-Salam (PS) maximal subgroup.

We now specify how the symmetry is broken down to the MSSM gauge
group \cite{aulmoh,ckn,abmsv} by superlarge vevs contained in the
 ${\bf{210, 126},{\overline{126}}}$-plet scalar vevs. Before doing so we introduce our submultiplet  naming and indexing
conventions. A host of further details related to the Pati Salam
decomposition of SO(10) can be found in our earlier paper
\cite{alaps} where the foundation for the current program of
computation of states, masses and couplings of this theory was
laid and the spectrum of MSSM like $SU(2)_L $ doublets and
$SU(3)_c$ ``baryon decay'' triplets first computed.

We denote  quantum numbers w.r.t the SM gauge group by enclosing
them in square brackets  while those with respect to the PS group
are denoted by round brackets. We have adopted the rule that any
PS submultiplet of an SO(10) field is always denoted by the
{\it{same}} symbol as its parent field, its identity being
established by the indices it carries or by additional
sub/superscripts ($(a),(s),\pm,L,R$)
 denoting (anti-)symmetry or (anti)-self duality, if
necessary. On the other hand, since one often  encounters several
chiral MSSM multiplets of the same type arising from different
SO(10) Higgs multiplets we will  also introduce a naming
convention using roman letters for these multiplets. If we need to
denote the scalar component of a chiral superfield
 we use a tilde over the superfield symbol and sometimes
use a superscript ``F'' to denote fermionic components of chiral
superfields while gauginos  are denoted by $\lambda $. Our
notation
 for indices is as follows : The real indices of the
vector representation of SO(10)
 are denoted by $i,j =1..10 $. The {\it {real}}
vector index of the upper left block embedding (i.e. the embedding
specified by the breakup of the vector multiplet $10=6 + 4$) of
SO(6) in SO(10) are denoted $a,b=1,2..6$ and of the lower right
block embedding of SO(4) in SO(10) by
${\tilde{\alpha},\tilde\beta= 7,8,9,10}$. These indices are
complexified via a Unitary transformation and denoted by
$\hat{a},\hat{b}=\hat{1},\hat{2},\hat{3},\hat{4},\hat{5},\hat{6}
\equiv \overline{\mu},\overline{\mu}^{*}=
\bar{1},{\bar{1}^*,\bar{2},\bar{2}^*,\bar{3},\bar{3}^*}$ where
$\hat{1}\equiv \bar{1}, \hat{2}\equiv \bar{1}^* $ etc. Similarly
we denote the complexified versions of
${\tilde{\alpha},\tilde\beta}$ by $\hat{\alpha},\hat{\beta}=
\hat{7},\hat{8},\hat{9},\widehat{10}\equiv {\hat 0}$.
 Using this complexification we
showed\cite{alaps}
 how all $SO(6) \times SO(4)$ subinvariants of SO(10)
tensor invariants could be systematically converted to \PS
 invariants whose indices are as follows :
The indices of the doublet of SU(2)$_L(\rm{SU(2)}_R$) are denoted
$\alpha,\beta=1,2$($\dot\alpha,\dot\beta=\dot{1},\dot{2})$.
Finally the index of the fundamental {\bf{4}}-plet of SU(4) is
denoted by a (lower) $\mu,\nu = 1,2,3,4$ and its upper-left block
SU(3) subgroup indices are $\bar\mu,\bar\nu = 1,2,3$. The
corresponding indices on the ${\bar{\bf 4}}$ are carried as
superscripts. These doublets and quartets correspond to the chiral
spinor representations of the $SO(4)$ and $SO(6)$ subgroups of
$SO(10)$. Details of the spinorial invariant decomposition
techniques may be found in \cite{alaps}. The component of the
SU(4) adjoint in the direction of the Gell-Mann generator
${i\lambda^{(15)}\over{\sqrt{2}}}$ is labelled with a superscript
$(15)$ or $(B-L)$.

Thus the PS decomposition of our SO(10) multiplets is \bea \phi =
210 &=& \phi_{\mu}^{~\nu}(15,1,1) + \phi(1,1,1) +
\vec\phi_{\mu(L)}^{~\nu} (15,3,1)+\vec\phi_{\mu(R)}^{~\nu}(15,1,3)
\nonumber \\ &+&  \phi_{{\mu\nu} ,\alpha\dot\alpha(a)}(6,2,2) +
\phi_{\mu\nu,\alpha\dot\alpha(s)} (10,2,2) +
\overline{\phi}^{\mu\nu}_{~~\alpha\dot\alpha(s)}(
\overline{10},2,2) \eea
\be \Sigma=\Sigma^{+} = 126 = \vec\Sigma^{\mu\nu(s)}_{R}
(\overline{10},1,3)+ \vec\Sigma_{\mu\nu(s)}^{L} (10,3,1) +
\Sigma^{\mu\nu}_{(a)}(6,1,1) + \Sigma_{\mu}^
{~\nu,\alpha\dot\alpha}(15,2,2) \ee
\be \overline{\Sigma}=\Sigma^{-} = \overline{126} =
{\vec{{\overline\Sigma}}^{R(s)}}_ {\mu\nu} (10,1,3) +
{\vec{{\overline\Sigma}}^{\mu\nu(s)}}_{L} (\overline{10},3,1) +
\overline{\Sigma}^ {\mu\nu}_{(a)}(6,1,1) +
\overline{\Sigma}_{\mu}^{~\nu\alpha\dot\alpha}(15,2,2) \ee
\be H = 10 = H_{\alpha\dot\alpha} (1,2,2) + H_{\mu\nu}(6,1,1) \ee
\be
 \Psi_A=16=16_{+}=(4_{+},2_{+})+(4_{-},2_{-})=
{\bar F^{\mu}}_{\da}(\overline{4},1,2)+F_{\mu\alpha}(4,2,1) \ee

\bea  F(4,2,1) = (Q_{\bm\alpha},L_{\alpha} )
 \qquad \qquad {\bar F}({\overline{4}},1,2) =
 ({\overline{Q}}_{\dot\alpha}^{\bm},{\overline{L}}_{\alpha}) \eea

with
\begin{equation}
Q=\left({\begin{array}{c}U\\D\end{array}}\right)
 \quad L=\left({\begin{array}{c} \nu\\e\end{array}}\right) \quad
{\overline Q}=\left({\begin{array}{c} {\bar d}\\{\bar
u}\end{array}}\right) \quad
{\overline{L}}=\left({\begin{array}{c}{\bar e}\\{\bar
\nu}\end{array}}\right) \quad
\end{equation}

The GUT scale vevs that break the gauge symmetry down to the SM
symmetry are {\cite{aulmoh,ckn}}:

 \be
{\langle(15,1,1)\rangle}_{210}:\langle{\phi_{abcd}}\rangle={a\over{2}}
\epsilon_{abcdef}\epsilon_{ef} \ee

   where~
$[\epsilon_{ef}]=Diag(\epsilon_{2},\epsilon_{2},\epsilon_{2}),~~\epsilon_{2}=i\tau_{2}
$. One dualizes \be
\phi_{ab}\equiv{1\over{4!}}\epsilon_{abcdef}\phi_{cdef}\ee

 Then in SU(4) notation $[\phi_{\nu}^{~\lambda}]$  this vev is
\be [\langle{\phi_{\nu}^{~\lambda}}\rangle]={ia
\over{2}}Diag(I_{3},-3)\equiv {ia\Lambda \over {2}} \ee

\be
\langle(15,1,3)\rangle_{210}~:~\langle\phi_{ab\ta\tb}\rangle=\omega
\epsilon_{ab}\epsilon_{\ta\tb} \ee
 where $[\epsilon_{\ta\tb}]=Diag(\epsilon_{2},\epsilon_{2})$
 which translates to
\be \langle(\vec{\phi}^{(R)\nu}_{\mu~\dot{1}\dot{2}})\rangle=
-{\omega\Lambda\over \sqrt{2}}\equiv
{i\langle(\vec{\phi}^{(R)\nu}_{\mu})_{0}\rangle} \ee

 \be
\langle(1,1,1)\rangle_{210}~: ~\langle\phi_{ {\tilde
\alpha}{\tilde \beta} {\tilde \gamma}{\tilde \delta}}
\rangle=p\epsilon_{{\tilde \alpha} {\tilde \beta} {\tilde
\gamma}{\tilde \delta}}
 \ee

  \be
 \langle(10,1,3)\rangle_{\overline{126}}~:~\langle\overline{\Sigma}_{\hat{1}\hat{3}\hat{5}
\hat{8}\hat{0}}\rangle=
\bar\sigma=-i\langle\overline{\Sigma}^{(R)}_{44(+)}\rangle={\overline{\Sigma}_{44\dot{1}\dot{1}}
\over{\sqrt{2}}} \ee
 \be
\langle(\overline{10},1,3)\rangle_{{126}}~:~\langle{\Sigma}_{\hat{2}\hat{4}\hat{6}\hat{7}\hat{9}}\rangle=
\sigma=i
\langle{\Sigma}^{(R)44}_{(-)}\rangle={\Sigma^{44}_{\dot{2}\dot{2}}
\over{\sqrt{2}}}\ee
%

Substituting these vevs into the superpotential one obtains

 \bea
W&=&m(p^2+3a^2+6\omega^2)+2\lambda({a^{3}}+3p\omega^{2} +6 {a}
\omega^2) \nonumber\\ &+& (M + \eta(p+ 3 a - 6\omega))
\sigma{\overline\sigma}
 \eea

the nontrivial F term conditions are thus :

 \bea
 2 mp+6\lambda\omega^2+\eta\sigma\bar\sigma&=&0\\
 2 ma+2\lambda(a^2+2\omega^2)+\eta\sigma\bar\sigma&=&0\\
 2 m\omega+2 \omega\lambda(p+2a)-\eta\sigma\bar\sigma&=&0\\
 (M+\eta(p+3a-6\omega)) \sigma&=&0
\eea

The vanishing of the D-terms of the SO(10) gauge sector
 potential imposes
only the condition \be |\sigma|=|{\overline{\sigma}}| \ee

Except for  degenerate cases corresponding to enhanced unbroken
symmetry\hfil\break ($SU(5)\times U(1), SU(5), G_{3,2,2,B-L},
G_{3,2,R,B-L}$ etc)\cite{abmsv,bmsv} this system  of equations is
essentially cubic and can be reduced to  the single cubic equation
\cite{abmsv}
 for a variable $x= -\lambda\omega/m$ :

\be 8 x^3 - 15 x^2 + 14 x -3 = -\xi (1-x)^2 \label{cubic} \ee

where  $\xi ={{ \lambda M}\over {\eta m}} $ and the other vevs can
be expressed in terms of values  of the variable $x$ which solve
eqn(\ref{cubic}).This parametrization of the MSGUT ssb problem
\cite{abmsv} is of great help computationally and clearly exhibits
the crucial importance of the $\xi$ i.e of the ratio $M/m$. The
important role played by a similar
 ratio in the other renormalizable SO(10) GUT based on the
 ${\bf{45,54,126,\oot}}$ representations
  has already been noted \cite{abmrs}.

When we  measure vevs or masses in units of ${\tilde m}=m/\lambda$
we will put a tilde over the symbol. We also define the additional
dimensionless parameters $\et=\eta/\lambda$ and
${\mht}=M_H/{\tilde m}$.

Then the dimensionless vevs are $\omt=-x$  \cite{abmsv} and

\be \at={{ (x^2 +2 x -1)}\over (1-x)}\quad ;\quad
 \pt={{x(5 x^2-1)}\over {(1-x)^2}}\quad ; \quad
\sgt\sgbt={2\over \et}{{x (1-3x)(1+x^2)}\over {(1-x)^2}}
\label{dlvevs}\ee

The solutions of the cubic equation (\ref{cubic}) are generically
complex. We will therefore nowhere  assume hermiticity for our
mass matrices, preferring to leave them undiagonalized for
eventual numerical diagonalization
  so that all our results are
applicable in the general case. We will not generate arrays of
expressions in terms of the variable x , although it is easy  to
do so since, practically speaking, the substitutions are now
handled via a computer anyway.

We conclude this section with a description of the super-Higgs
effect for the breaking $SO(10)\longrightarrow SU(3)\times SU(2)
\times U(1)_Y $
  which is achieved by the above
superheavy vevs. As is well known,
 as a consequence of gauge symmetry breaking,
  each massive gauge boson forms a massive supermultiplet
together with its longitudinal goldstone pseudo scalar (and its
real scalar partner) as the 4 bosonic degrees of freedom. It's
gaugino and the
  chiral fermion superpartner of the
Goldstone scalar pair  make up one Dirac fermion super-partner
also with 4 degrees of freedom.
 This is  the so called Dirac or massive vector gauge
supermultiplet. These gauge boson/gaugino masses are the most
fundamental thresholds of the GUT and it is appropriate to begin
with a discussion of their values for this model.
 In Section 4. we follow \cite{weinberg,hall} to
 compute the threshold corrections using the spectra we compute.
 Then a Dirac gauge coset multiplet  in representation R
of the MSSM  gives rise to a  RG mass threshold above which the
gauge and chiral components of the Dirac multiplets separately
contribute $-3T_i(R)$ and $T_i(R)$, respectively, to the beta
function coefficients of the individual MSSM couplings (including
the $U(1)_Y$ coupling !).

 It is easiest to keep track of the gaugino masses and mixings.
The combination of chiral fermions that forms a Dirac fermion
together with a gaugino must, for consistency, be a zero mode of
the mass matrix arising from the superpotential and this makes it
easy to disentangle the gauge spectrum even in the case of complex
vevs and parameters. For the symmetry breaking to the MSSM the
gauginos of the coset $SO(10)/G_{321}$ lying in the PS
representations $(6,2,2) \oplus (1,1,3)$ plus  the triplets and
anti-triplets in $(15,1,1)$ (i.e 33 Dirac multiplets in all)
obtain a mass by pairing with chiral AM Higgs fermions.
   One need   only substitute the vevs given above
into the PS decomposition of the gaugino
 Yukawa terms which have the  form

\be g {\sqrt 2} \big\{ {1\over{3!}} <{\tilde\Phi}_{ijkl}^*>
\lambda_{im} \Phi^F_{mjkl} + {1\over{2.4!}}(
<{\tilde\Sigma}_{ijkln}^*>
 \lambda_{im} \Sigma^F_{mjkln} +
<{\tilde{\overline \Sigma}}_{ijkln}^*> \lambda_{im}
{\overline\Sigma}^F_{mjkln}) \big\} + H.c \ee

One finds the following gaugino masses :
\begin{itemize}

\item{(i)} G[1,1,0] : $m_{\lambda_G}={\sqrt {10}} g |\sigma| $.

The mass term is

\bea {g\over{{\sqrt 2}}} ({{\sqrt 2}}\lambda^{({R0})} &-&{{\sqrt
3}} \lambda^{(15)}) (\sigma^* \Sigma^{44}_{R-}
  +  {\bar\sigma}^* {\overline\Sigma}_{44R+}) + H.c \nonumber\\
 &\equiv&  m_{\lambda_G} {{G_6}\over{\sqrt 2}}
(e^{-i\gamma_{\sigma}} G_4 + e^{-i\gamma_{\bar\sigma}} G_5)  +
H.c. \nonumber \\
G_6 &\equiv& ({\sqrt {2\over 5}}\lambda^{({R0})}
  - {\sqrt {3\over 5}} \lambda^{(15)})\nonumber \\
G_4 &\equiv&{{\Sigma^{44}_{R-}}\over{\sqrt 2}} \quad;\quad
 G_5 = {{\Sigb_{44}^{R+}}\over{\sqrt 2}}
 \eea
The naming conventions for the chiral states are given in Section
IV and the Appendix . Here $\gamma_{\sigma},\gamma_{\bar\sigma}$
are the phases of $\sigma, {\bar \sigma}$. Since the
representation is real, the mass matrix ${\cal G}$ in this sector
is symmetric. The complete G$[1,1,0]$ sector mass
matrix(includiing gauginos)${\cal G}$ is $6\times 6$ while its
pure chiral part {\bf{G}} (which  arises only from the
superpotential)  is $5 \times 5$ and symmetric and the 5-tuple
$(0,0,0, \sigma,{\bar\sigma}) $ is both a left and right null
eigenvector of {\bf{G}}  - as will be obvious when it is presented
further on (Section 2. and Appendix).

\item{ii)}${\bar J}[{\bar 3},1,-{4\over 3}] \oplus J [3,1,{4\over
3}]   :\qquad\qquad m_{\lambda_J}=g {\sqrt{ 8 |a|^2 + 16|\omega|^2
+ 2 |\sigma|^2}} $

In this case $(J_4)_{\bm}= \lambda_{\bm}^{~ 4}$,
 $ ({\bar J}_4)^{~\bm}=  \lambda_4^{~\bm } $ pair up
with the combinations corresponding to the left and right null
eigenvectors $v_{0JL}=N_J( -\bar\sigma,2 a, 2 {\sqrt 2} \omega  )
, \quad v_{0JR}^T= N_J (  \sigma ,2 a, 2 {\sqrt 2} \omega  ) $ of
the complex,
 non symmetric, upper left $3\times 3$ sub-matrix ${\bf J} $
of the  $4\times 4$  mass matrix ${\cal J}$ in the J sector. The
gaugino mass terms are \be ig{\bar J}_4 (2\sq a^* J_2 + 4 \om^*
J_3 -\sq\ssb^* J_1) -ig(2\sq a^* {\bar J}_2 + 4 \om^* {\bar J}_3
+\sq\sss^* {\bar J}_1){J}_4  + H.c \ee

 \item{iii)} ${\bar F}[1,1,-2] \oplus    F[1,1,2] \qquad ;
\qquad\qquad m_{\lambda_F}= g {\sqrt{24 |\omega|^2 +2
|\sigma|^2}}$

The chiral partners of the gauginos $F_3 \equiv \lambda_{R+} ,
{\bar F}_3\equiv\lambda_{R-} $ correspond to the right and left
 null eigenvectors $v_{0FR}=(-\sigma,
 {\sqrt{12}}i\omega)^T ;\quad
v_{0FL}=(\bar\sigma, {\sqrt{12}}i\omega) $ of the $2\times 2$
 $ \bar F-F$  chiral fermion mass matrix. The mass terms are

\be -g {\bar F}_3 ( i \om^*{\sqrt{24}} F_2 + \sq \sss^* F_1) + g
(i\om^* {\sqrt{24}} {\bar F}_2 - \sq \ssb^* {\bar F}_1)F_3 \ee

\item{iv)} ${\bar E}[\bar 3, 2,-{1\over 3}]
 \oplus  E[3,2,{1\over 3}]
\quad ; $ $\qquad m_{\lambda_E}= g{\sqrt{(4|a-\omega|^2 + 2|w-p|^2
+ 2|\sigma|^2)}} $.

The  chiral partners of the gauginos $E_5 \equiv \lambda_{\bm  4
{\dot 1}\alpha} , {\bar E}_5 \equiv \lambda^{\bm 4}_{{\dot
2}\alpha} $
 correspond to the  null eigenvectors
$v_{0ER}=(i\sigma, {\sqrt{2}}(a-\omega),\omega-p)^T  ; \qquad
v_{0EL}=(-i\bar\sigma, {\sqrt{2}}(a-\omega), \omega-p) $
 of the upper left  $3\times 3$ corner ${\bf{E}}$
  of the  $ E$ sector  $4 \times 4$ chiral fermion mass
 matrix ${\cal E}$. $E_1,{\bf{\bar E_1}}$ do not mix with other E-sector multiplets.
  The mass terms are

\bea
&& g{\bar E}_5 (-i\sq \sss^* E_2 + 2(a^*-\om^*)E_3 +\sq(\om^*-p^*) E_4 ) \nonumber \\
&+&g ((i\sq \ssb^* {\bar E}_2 + 2(a^*-\om^*){\bar E}_3
 +\sq(\om^*-p^*) {\bar E}_4 ) {E}_5
 \eea

{\item{v)}}
 ${\overline X}[\overline 3,2,{5\over 3}]\oplus     X[3,2,{-{5\over 3}}] $ :
$\quad m_{\lambda_X}= g\sqrt{4 |a+\omega|^2 + 2 |p+\omega|^2 }$.

The  chiral partners of the gauginos $X_3 \equiv \lambda_{\bm 4
{\dot 2}\alpha} , {\overline X}_3 \equiv \lambda^{\bm 4}_{{\dot
1}\alpha} $
 correspond to the  null eigenvectors
$v_{0XR}=(-{\sq}(a+\omega), \omega+p)^T =v_{0XL}   $
 of the upper left $2\times 2$ corner ${\bf{X}} $
  of the $3\times 3 $  X-sector
chiral fermion mass matrix $\cal X$. The X-gaugino mass terms are
:

\be g{\overline X}_3 (-2(a^* + \om^*) X_1 + \sq(p^* +\om^*) X_2) +
g (-2(a^* + \om^*) {\overline X}_1 + \sq(p^* +\om^*) {\overline
X}_2) X_3 \ee
 \end{itemize}

\section{  AM Chiral masses via PS}

Our approach to opening up the maze of MSSM interactions coded in
the deceptive simplicity
 of the SO(10) form of the GUT action is to
rewrite SO(10) invariants as combinations of PS invariants using
the translation techniques developed by us \cite{alaps}. Although
tedious, our  approach allows one to keep track of all phases and
normalizations without any ambiguity. Once this is done making
contact with the MSSM phenomenology becomes trivial since the
embedding \GSM $\subset$  \PS  is  trivial and transparent if one
keeps in mind that

$$Y=2T_{3R} + B-L $$

 We obtain for the PS form of the different terms in $W_{AM}$

\be
{m\over4!}\phi^{2}_{ijkl}=-m\{\phi_{\mu}^{~\nu}\phi_{\nu}^{~\mu}+
\phi_{\mu\nu(s)}^ {~\alpha\dot\alpha}\bar{\phi}^{\mu\nu(s)}_
{\alpha\dot\alpha}+
\vec{\phi}_{\mu(R)}^{~\nu}.\vec{\phi}_{\nu(R)}^{~\mu}
+\vec{\phi}_{\mu(L)}^{~\nu}.\vec{\phi}_{\nu(L)}^{~\mu} +
{1\over2}\tilde{\phi}^{\mu\nu\alpha\dot\alpha}_{(a)}
\phi_{\mu\nu\alpha\dot\alpha}^{(a)}-\phi^2\} \ee
\be {M \over 5!}\Sigma_{ijklm}\overline{\Sigma}_{ijklm}
=M\{\widetilde\Sigma^{\mu\nu}_{(a)}
\overline{\Sigma}_{\mu\nu}^{(a)}+2{\Sigma^
{~~\mu}_{\nu}}^{\alpha\dot\alpha}{\overline{\Sigma}_{\mu}^{~\nu}}_{\alpha\dot\alpha}
+(\vec{\os}_{\mu\nu(R)}^{(s)}.{\vec\s}^{\mu\nu(s)}_{(R)}+
\vec{\s}_{\mu\nu(L)}^{(s)}. {\vec{\os}^{\mu\nu}}_{(s)(L)})\} \ee
\bea
 {1 \over 4!}\lambda\phi^{3}&=&\lambda[-{2\over 3}i\phi_{\mu}^{~\nu}
\phi_{\nu}^{~\lambda}\phi_{\lambda}^{~\mu}-2i(\phi^{~\nu}_{\mu}\phi
_{\nu\lambda(s)}^{\alpha\dot\alpha}\bar{\phi}^{\lambda\mu(s)}_
{~\alpha\dot\alpha}) \nonumber\\
 &-&2i\{\phi_{\mu}^{~\nu}(\vec{\phi}_{\nu(R)}^
{~\lambda}.\vec{\phi}_{\lambda(R)}^{~\mu}
+\vec{\phi}_{\nu(L)}^{~\lambda}.\vec{\phi}_{\lambda(L)}^{~\mu})\}
\nonumber\\
&+&\{(\tilde{\phi}^{\mu\nu\alpha\dot\alpha}_{(a)}(\phi_{\mu\lambda(s)
\alpha}^{\dot\beta}\phi_{\nu(R)\da\db}^{~\lambda}-\phi_{\mu\lambda(s)\da}
^{\beta}\phi_{\nu(L)\alpha\beta}^{~\lambda}))
 \nonumber\\
&+&(\phi_{\mu\nu(a)}^{\alpha\dot\alpha}(-\phi^{\mu\lambda~\db}
_{\alpha}\phi_{\lambda(R)\da\db}^{~\nu}+\phi^{\mu\lambda(s)\beta}_{\da}
\phi_{\lambda(L)\alpha\beta}^{~\nu}))\}\nonumber\\
&+&\sqrt{2}\{\phi_{\nu\lambda(s)}^{~\alpha\da}({\phi^{\lambda\mu}_{(s)}}
_{\alpha}^{~\db}\phi_{\mu(R)\da\db}^{~\nu}+{\phi^{\lambda\mu}_{(s)}}^{\beta}_
{\da}\phi_{\mu(L)\alpha\beta}^{~\nu})\} \nonumber\\
&-& \phi\{(\vec{\phi}_{\nu(R)}^
{~\lambda}.\vec{\phi}_{\lambda(R)}^{~\nu}
-\vec{\phi}_{\nu(L)}^{~\lambda}.\vec{\phi}_{\lambda(L)}^{~\nu})\}\nonumber\\
&-&{1\over{\sqrt
2}}\{\phi_{\mu\nu{(a)}}^{~\alpha\da}({\phi^{\nu\lambda}
_{(a)}}_{\alpha}^{\db}\phi_{\lambda(R)\da\db}^{~\mu}+{\phi^{\nu\lambda}
_{(a)}}_{\da}^{\beta}\phi_{\lambda(L)\alpha\beta}^{~\mu})\} \nonumber\\
&+&{\sqrt{2}\over{3}}\{\phi_{\mu(R)}^{~\nu~\da\db}{\phi_{\nu}^
{~\lambda}}_{(R)\db}^{~\dot\gamma}\phi_{\lambda(R)\dot\gamma\da}^{~\mu}
+\phi_{\mu(L)}^{~\nu\alpha\beta}{\phi_{\nu}^
{~\lambda}}_{(L)\beta}^{~\gamma}
\phi_{\lambda(L)\gamma\alpha}^{~\mu}\}] \eea
\bea {1\over 4!}\bar\gamma\phi{\overline\Sigma}H&=&\bar\gamma [i
H_{\mu\nu(a)}
{\overline{\Sigma}}^{\nu\lambda(a)}\phi_{\lambda}^{~\mu}+ \phi_
{\mu}^{~\nu}{\overline{\Sigma}}_{\nu}^{~\mu\alpha\da}H_{\alpha\da}
 \nonumber\\
&-&{1\over 2}H^{\alpha\da}(\phi^{\mu\nu~\db}_{(s)\alpha}
{\overline{\Sigma}}_{\mu\nu(s)\da
\db(R)}+\phi_{\mu\nu~\da}^{(s)\beta}
{\overline{\Sigma}}^{\mu\nu(s)}_{\alpha\beta(L)})\}\nonumber\\
&-&{1\over\sqrt{2}}\{\tilde{H}^{\mu\nu(a)}{\overline{\Sigma}}_{\nu}
^{\lambda\alpha\da}\phi_{\mu\lambda(s)\alpha\da}+H_{\mu\nu(a)}
{\overline{\Sigma}}_{\lambda}^{~\nu\alpha\da}\bar\phi^{\mu\lambda(s)}
_{\alpha\da}\}
 \nonumber\\
&-&{i\over\sqrt{2}}\{H^{\alpha\da}({\overline{{\Sigma}}_{\mu~\alpha}
^{~\nu~\db}}
\phi_{\nu(R)\da\db}^{~\mu}-{{\overline{\Sigma}}_{\mu~\da}^{~\nu~\beta}}
\phi_{\nu(L)\alpha\beta}^{~\mu})\}\nonumber\\
&+&(-\tilde{H}^{\mu\nu(a)}\vec\phi_{\nu(R)}^{~\lambda}.
\vec{\overline{\Sigma}}_{\mu\lambda(s)(R)}+H_{\mu\nu(a)}
\vec\phi_{\lambda(L)}^{~\nu}.\vec{\overline{\Sigma}}^{\mu\lambda(s)}_{(L)})
 \nonumber\\
&-&{1\over{2}}\widetilde{\overline{\Sigma}}^{\mu\nu(a)}\phi
^{~\alpha\da}_{\mu\nu(a)}H_{\alpha\da}\nonumber\\
&-& i
H_{\mu\nu(a)}\phi^{\nu\lambda(a)\alpha\da}{{\overline{\Sigma}}_
{\lambda}^{~\mu}}_{\alpha\da}+{1
\over2}\phi\widetilde{H}^{\mu\nu(a)}{\overline{\Sigma}}_{\mu\nu(a)}]
\label{phSbH}\eea
\bea {1\over 4!}\gamma\phi\Sigma H&=&\gamma[-i H_{\mu\nu(a)}
{\Sigma}^{\nu\lambda(a)}\phi_{\lambda}^{~\mu} + \phi_
{\mu}^{~\nu}{\Sigma}_{\nu}^{~\mu\alpha\da}H_{\alpha\da}
 \nonumber\\
&-&{1\over 2}H^{\alpha\da}(\phi_{\mu\nu~\alpha}^{(s)\db}{\Sigma}^
{\mu\nu(s)}_{\da \db(R)}+\phi^{\mu\nu~\beta}_{\da}
{\Sigma}_{\mu\nu(s)\alpha\beta(L)})\nonumber\\
&-&{1\over\sqrt{2}}\{\tilde{H}^{\mu\nu(a)}\Sigma_{\nu}
^{\lambda\alpha\da}\phi_{\mu\lambda(s)\alpha\da}+H_{\mu\nu(a)}
\Sigma_{\lambda}^{~\nu\alpha\da}\bar\phi^{\mu\lambda(s)}
_{\alpha\da}\} \nonumber\\
&+&{i\over\sqrt{2}}\{H^{\alpha\da}({{\Sigma}_{\mu~\alpha}
^{~\nu~\db}}
\phi_{\nu(R)\da\db}^{~\mu}-{{\Sigma}_{\mu~\da}^{~\nu~\beta}}
\phi_{\nu(L)\alpha\beta}^{~\mu})\}\nonumber\\
&+&(-\tilde{H}^{\mu\nu(a)}\vec\phi_{\nu(L)}^{~\lambda}.
\vec{\Sigma}_{\mu\lambda}^{(s)(L)}+H_{\mu\nu}^{(a)}
\vec\phi_{\lambda(R)}^{~\nu}.\vec{\Sigma}_{\mu\lambda}^{(s)(R)})
 \nonumber\\
&-&{1\over{2}}\widetilde{\Sigma}^{\mu\nu(a)}\phi^{~\alpha\da}_
{\mu\nu(a)}H_{\alpha\da}\nonumber\\
&+& i H_{\mu\nu(a)}\phi_{(a)}^{\nu\lambda\alpha\da}{{\Sigma}_
{\lambda}^{~\mu}}_{\alpha\da}+{1
\over2}\phi\widetilde{H}^{\mu\nu}_{(a)}{\Sigma}_{\mu\nu}^{(a)}\}
\label{phSH}\eea
\bea {\eta \over{4!}}\phi\Sigma{\overline{\Sigma}}&=& \eta
[2i\phi_{\mu}^{~\nu}(\s_{\nu}
^{~\lambda\alpha\da}\os_{\lambda\alpha\da}^{~\mu}+\os_{\nu}^
{~\lambda\alpha\da}\s_{\lambda~\alpha\da}^{~\mu}) \nonumber\\
&+&2i\phi_{\mu}^{~\nu}(\vec{\os}_{\nu\lambda(s)(R)}.\vec{\s}_{(s)(R)}^
{\mu\lambda}+\vec{\s}_{\nu\lambda(s)(L)}.\vec{\os}_{(s)(L)}^{\mu\lambda})
\nonumber\\
&+& \phi(\vec{\os}_{\mu\nu(s)(R)}.\vec{\s}_{(s)(R)}^
{\mu\nu}-\vec{\s}_{\mu\nu(s)(L)}\cdot{\vec{\os}}_{(s)(L)}^{\mu\nu})
\nonumber\\
&+&i\sqrt{2}(-\tilde{\s}^{\mu\nu(a)}\os_{\nu}^{~\lambda~\alpha\da}
\phi_{\mu\lambda(s)\alpha\da}+\s_{\mu\nu(a)}\os_{\lambda}^{~\nu~\alpha\da}
\phi^{\mu\lambda(s)}_{\alpha\da}) \nonumber\\
&+&i\sqrt{2}(\tilde{\os}^{\mu\nu(a)}\s_{\nu}^{~\lambda~\alpha\da}
\phi_{\mu\lambda(s)\alpha\da} -
\os_{\mu\nu(a)}\s_{\lambda}^{~\nu~\alpha\da}
\phi^{\mu\lambda(s)}_{\alpha\da})\nonumber\\
&-&2i\s^{~\nu\alpha\da}_{\mu}(\phi_{\nu\lambda(s)\da}^{~\beta}\os^
{\mu\lambda(s)}_{\alpha\beta(L)}+\os_{\nu\lambda\da\db(R)}^{(s)}
\phi^{\mu\lambda(s)\db}
_{\alpha}) \nonumber\\
&-&2i\os_{\mu}^{~\nu\alpha\da}(\phi_{\nu\lambda(s)\alpha}^{~\db}\s^
{\mu\lambda(s)}_{\da\db(R)}+\s_{\nu\lambda\alpha\beta(L)}^{(s)}\phi^
{\mu\lambda(s)\beta}_{\da})\nonumber\\
&-&2(\tilde\s^{\mu\nu(a)}\vec{\phi}_{\nu(R)}^{\lambda}.\vec{\os}_
{\mu\lambda(R)}+\s_{\mu\nu(a)}\vec{\phi}_{\lambda(L)}^{\nu}.\vec{\os}^
{\mu\lambda(L)}) \nonumber\\
&+&2(\tilde{\os}^{\mu\nu(a)}\vec{\phi}_{\nu
(L)}^{\lambda}.\vec{\s}_
{\mu\lambda(L)}+\os_{\mu\nu(a)}\vec{\phi}_{\lambda
(R)}^{\nu}.\vec{\s}^
{\mu\lambda(R)})\nonumber\\
&-&2\sqrt{2}(
\os_{\mu}^{~\nu\alpha\da}\s_{\nu\alpha}^{~\lambda\db}\phi
_{\lambda(R)\da\db}^{~\mu}+ \s_{\mu}^{~\nu\alpha\da}
\os_{\nu\da}^{~\lambda\beta}
 \phi_{\lambda(L)\alpha\beta}^{\mu} )  \nonumber\\
&-&\sqrt{2}(\phi_{\nu~(R)}^{~\mu~\da\db}\os_{\mu\lambda\db}
^{(s)(R)\dot\gamma}
\s_{(R)\dot\gamma\da}^{\nu\lambda(s)}+\phi_{\nu~(L)}^{~\mu~\alpha\beta}
\s_{\mu\lambda\beta}^{(s)(L)\gamma}
\os_{(L)\gamma\alpha}^{\nu\lambda(s)})\nonumber\\
&+&i\sqrt{2}(-\tilde\phi^{\mu\nu\alpha\da}_{(a)}\os_{\nu~\da}
^{~\lambda~\beta}\s_{\mu\lambda\alpha\beta}^{(s)(L)}+\phi_{\mu\nu(a)}^
{\alpha\da}\os_{\lambda~\alpha}^{~\nu~\db}\s^{\mu\lambda(s)}_{\da\db(R)})
\nonumber\\
&+&i\sqrt{2}(\tilde\phi^{\mu\nu\alpha\da}_{(a)}\s_{\nu~\alpha}
^{~\lambda~\db}\os_{\mu\lambda\da\db}^{(s)(R)} -\phi_{\mu\nu(a)}^
{~\alpha\da}\s_{\lambda~\da}^{~\nu~\beta}
\os^{\mu\lambda(s)}_{\da\db(L)})]\label{phSSb}\eea

The purely chiral superheavy supermultiplet masses can be
determined from these expressions simply by substituting in the AM
Higgs vevs and breaking up the contributions
 according to MSSM labels.

It is again easiest to keep track of Chiral fermion masses since
all others follow using supersymmtery and the organization
provided by the gauge super Higgs effect.

There are three types of   mass terms involving fermions from
chiral supermultiplets in such models : (A) Unmixed Chiral (B)
Mixed pure chiral (C) Mixed chiral and gaugino .

\subsection{Unmixed Chiral}

A pair of Chiral fermions transforming as \GSM conjugates pairs up
to form a massive Dirac fermion . For example for the
 properly
normalized fields \be {\bar A}[1,1,-4] ={ {\overline \Sigma}_{44
(R-)}\over {\sqrt 2}} \qquad { A}[1,1,4] = {
{\Sigma}^{44}_{(R+)}\over {\sqrt 2}} \ee
 one obtains the mass term

$$ 2(M + \eta (p + 3a + 6 \omega)) {\bar A} A = m_A  {\bar A} A $$
The physical Dirac fermion mass is then $|m_A|$ since the phase
can be absorbed by a field redefinition. By supersymmetry this
mass is shared by a pair of complex scalar fields with the same
quantum numbers. If the representation is real rather than complex
one obtains an extra factor of 2 in the   masses .
 There are in fact 19 types of
such multiplets and their (roman letter) lables are given along
with their masses and SO(10) origins in Table I in the Appendix.
The case of the sectors C$[8,2,\pm 1]$ and D$[3,2,\pm 1]$ bears
special mention. The mass terms for these  multiplets arise only
between pairs drawn one each from $\Sig(15,2,2),\Sigb(15,2,2)$ and
 there is no mixing between a $C,\bar C$ or $D,\bar D$
 drawn from the same SO(10) multiplet simply because the
 superpotential does not contain any term containing
 $\Sig^2$  or ${\bf{\Sigb}}^2$. This was the reason
 for the discrepancy in this sector between
 the results of\cite{alaps,bmsv} and \cite{fm}: there simply
 is no such mixing.

\subsection{Mixed Pure Chiral}

In this case there are no contributions from the gaugino Yukawas
or the D-terms to the supermultiplet masses, but there is a mixing
among several multiplets of the same SM quantum numbers. There are
only three such multiplet types :
\begin{itemize}
\item{a)} $ [8,1,0](R_1,R_2)\equiv (\phi_{\bm}^{~\bn},
\phi_{\bm(R0)}^{~\bn})$

These mix with mass matrix

\be \cal{R} = 2\left({\begin{array}{cc} (m-\lambda a ) &
-\sqrt{2}\lambda\omega \\ -\sqrt{2}\lambda\omega & m+\lambda( p-a)
\end{array}}\right)
\ee with both rows and columns labelled by $(R_1,R_2)$. The masses
are  the magnitudes of the eigenvalues of the matrix
$\cal R$ .

\be |{\cal R}_{\pm} |= 2|  m [1 +({\pt \over 2} - \at) \pm
\sqrt{({{\pt}^2\over 4}) + 2 \omt^2}]|= {{m_{R_{\pm}}} }
 \ee

   While the corresponding eigenvectors can be found by
diagonalizing the matrix ${\cal R}{\cal R}^{\dagger}$.

 The  mass matrices of the electroweak doublets $h [1,2,1],
{\bar h} [1,2,-1]$
 and colour triplets $t [3,1,-{2\over 3}],
{\bar t}[\bar 3, 1,{2\over 3}]$ which mix with the   multiplets
contained in the {\bf {10}} plet FM Higgs are the most crucial
ones for determining the phenomenology of the effective MSSM that
arises from this GUT. These matrices were first calculated in
\cite{alaps}(v2) and later, stimulated by a contradiction with a
recent paper \cite{fm},
 the $d=5 $ baryon violating operators induced by
the exchange of heavy Higgsinos were computed and added to a
revised version \cite{alaps}(v4) by using the Clebsches for the $
16\cdot16\cdot\oot$  and $16\cdot 16\cdot10 $ invariants
calculated earlier by us. Thus one has :

\item{b)} $[1,2,-1]({\bar h}_1,\bh_2,\bh_3,\bh_4)  \oplus
[1,2,1](h_1,h_2,h_3,h_4)$\hfil\break $  \equiv  (H^{\alpha}_{
{\dot 2}},\s^{(15)\alpha}_{ \dot 2}, \Sigb^{(15)\alpha}_{\dot 2},
{{\phi_{44}^{\alpha\dot 2}} \over {\sqrt{2}}})  \oplus (H_{\alpha
{\dot 1}}, \Sigb^{(15)}_{\alpha \dot1}, \s^{(15)}_{\alpha\dot 1},
{{\phi^{44\dot 1}_{\alpha}} \over {\sq}}) $

These  multiplets label the 4  rows and columns   of the $4 \times
4 $ mass matrix $\cal H $\cite{alaps}
 which is given  in the
collection of mixing matrices in Appendix I . We note
  that we have redefined our mass parameters $m,M$ by a
factor of 2 relative to those we used in \cite{alaps}. To achieve
the MSSM spectrum of {\it{one pair}} of light doublets, it is
necessary to fine tune one of the  parameters of the
superpotential (e.g $M_H$) so that $Det{\cal H} =0$. By extracting
the null eigenvectors of ${\cal H}^{\dagger}{\cal H}$ and ${\cal
H}{\cal H}^{\dagger}$ one can compute the coefficients of the
various  bi-doublets in the light doublet pair, and, in
particular, we can find those for the doublets coming from the
${\bf 10, {\overline{126}}} $ multiplets which couple to the
matter sector (see Section 6.). In this way the SO(10) constraints
on the fitting of the Yukawa coupling matrices $h_{AB}',f_{AB}'$
can be brought into focus and the invalid assumption that the
squares of these coefficients\cite{babmoh,moh}
 add up to $1$  can be dispensed with.

 \item{(c)} $[\bar 3,1,{2\over 3}] (\bt_1,\bt_2,\bt_3,\bt_4,\bt_5)
\oplus [3,1,-{2\over 3}] (t_1,t_2,t_3,t_4,t_5)$\hfil\break $\equiv
(H^{\bm 4},\Sigb_{(a)}^{\bm 4}, \s^{\bm  4 }_{(a)},\s^{\bm
4}_{R0},\phi_{4(R+)}^{~\bm}) \oplus (H_{\bm 4},\Sigb^{(a)}_{\bm
4}, \s_{\bm 4 }^{(a)},\Sigb_{\bm 4}^{(R0)},\phi_{\bm(R-)}^{~4}) $

For generic values of the couplings all these particles are
superheavy. These triplets and antitriplets participate in baryon
violating process since the exchange of $(t_1,t_2,t_4)\oplus
(\bt_1,\bt_2) $ Higgsinos generates $d=5$ operators of type QQQL
and ${\bar l \bar u \bar u\bar d}$. The strength of the operator
is controlled by the inverse of the $\bt-t$ mass matrix $\cal T$
which we computed in \cite{alaps} and is given in the Appendix .
We shall examine how $d=5$ baryon and lepton number violating
operators are generated in Section 5..

\end{itemize}
\subsection{ Mixed Chiral-Gauge}

Finally we come to the mixing matrices for the chiral modes that
mix with the gauge particles as well as among themselves.Apart
from threshold effects, these are of some interest since one might
ask whether the new types of coset gauginos present in in SO(10)
but
 not in SU(5) namely $SO(10)/SU(5) \sim E[3,2,{1\over 3}]
\oplus{\bar E}[\bar 3,2,-{1\over 3}] \oplus F[1,1,2]\oplus
 {\bar F}[1,1,-2] \oplus  G[1,1,0]  \oplus J[3,1{4\over 3}]\oplus
{\bar J}[\bar 3,1,-{{4}\over 3}] $ (the $SU(5)/G_{321}$
leptoquarks are $X[3,2,{-{5\over 3}}] \oplus {\overline
X}[3,2,{5\over 3}]$ )   might not mediate interesting exotic
processes by inducing $d=5 $ operators via mixed gaugino-chiral
exchange. We have examined this question in some detail in Section
 5.

These multiplet sets are :
\begin{itemize}
\item{a)} $ [1,1,0] (G_1,G_2,G_3,G_4,G_5,G_6) \equiv
(\phi,\phi^{(15)},\phi^{(15)}_{(R0)},{{\s^{44}_{(R-)}}\over \sq},
{{\Sigb_{44(R+)}}\over \sq}, {{\sq \lambda^{(R0)} - {\sqrt{3}}
\lambda^{15)}\over {\sqrt{5}}}})$

which mix via a $6\times 6 $ mass matrix $\cal G$ given in the
Appendix. The complex conjugates of the 6th row and column form
left and right null eigenvectors  $v_{0GL}, v_{0GR}$ of the upper
left $5\times 5 $ block ${\bf{G}}$ of $\cal G$.
 The determinant of
 $\cal G$ is generically non zero although the determinant of the
submatrix ${\bf G} $ vanishes. It will clearly not affect the
evolution of the MSSM gauge couplings at one loop due to the
singlet quantum numbers .

\item{b)}  $[{\bar 3},2,-{1\over 3}]({\bar E}_2,{\bar E}_3,
 {\bar E}_4,{\bar E}_5) \oplus [3,2,{1\over 3}]
 (E_2,E_3,E_4,E_5) \equiv
\hfil\break ( \Sigb_{4\alpha\dot 1}^{~\bm},
\phi^{\bm4(s)}_{\alpha\dot 2}, \phi^{(a) \bm 4}_{\alpha\dot
2},\lambda^{\bm 4}_{\alpha\dot 2}) \oplus ( \s_{\bm\alpha\dot
2}^{~4} ,\phi_{\bm 4\alpha\dot 1}^{(s)}, \phi_{\bm 4\alpha\dot
1}^{(a)},\lambda_{\bm\alpha\dot 1}) $

The $4\times 4$ mass matrix $\cal E$ (($E_1,\bar E_1)\equiv
(\Sigb_{\bm\alpha\dot 2}^{~4},
  \s_{4\alpha\dot 1}^{~\bm}) $ do not mix with the
others) has the usual superhiggs structure : complex conjugates of
the 4th row and column are left and right null eigenvectors of the
upper left $3 \times 3 $ submatrix  ${\bf{ E}}$.
 $\cal E$ has non zero determinant
although the determinant of ${\bf{E}}$ vanishes. As for the case
of C$[8,2,\pm 1]$ and D$[3,2,\pm 7/3]$ type multiplets one finds
that the  conjugate types of E type multiplets drawn from the same
SO(10) representation cannot mix. Furthermore explicit computation
using the decomposition of the superpotential given in Section  3.
shows that $E_1[3,2,{1\over 3}]=\Sigb_{\bn\alpha\dot 2}^{~4}$ and
$\bar E_1[(\bar 3,2,- {1\over 3}]=  \s_{4\alpha\dot 1}^{~\bn}$ in
fact decouple from the other members of the E sector so that the E
sector mixing matrix is $4\times 4$ (including gauginos) and
$3\times 3$ excluding gauginos.
  Note that our assertion is not that  these couplings cancel but
simply that they do not appear.
 To see  why, for instance,  there is no term mixing say $E_1=\Sigb_{\bn\alpha\dot 2}^{~4}$  with
$E_3=\phi_{\bm 4\alpha\dot 1}^{(s)}$ coming from the
$\Phi(10,2,2)$ we observe that the terms mixing $\Sig(15,2,2)$ and
${\bf{\Phi}}(decuplet,2,2)$ via a righthanded vev could  only come
from the following two terms in   eqn.(\ref{phSSb}) :

\bea {\eta \over{4!}}\phi\Sigma{\overline{\Sigma}} =
-2i\s^{~\nu\alpha\da}_{\mu}( \os_{\nu\lambda\da\db(R)}^{(s)}
\phi^{\mu\lambda(s)\db} _{\alpha})  -
2i\os_{\mu}^{~\nu\alpha\da}(\phi_{\nu\lambda(s)\alpha}^{~\db}\s^
{\mu\lambda(s)}_{\da\db(R)} )\nonumber \eea
 We see that the pairs
 ${\bf{\Sigb}}(10,1,3)$ and ${\bf{\phi}}(10,2,2)$ and
 ${\bf{\Sig}}({\overline{10}},1,3)$ and ${\bf{\phi}}
 ({\overline{10}},2,2)$ simply do not mix. Now it is obvious that a $\Sig$ right
 handed vev will mix only $\bar E_2$ coming from $\Sigb(15,2,2)$ with $E_3$
 coming from $\Phi(10,2,2)$ {\it{but not}} $\bar E_1$ coming from
  $\Sig(15,2,2)$ with $E_3$  coming from $\Phi(10,2,2)$. Similar
  considerations account for the other decouplings between $E_1,\bar E_1$
  and the rest of the E sector.
  Ultimately this correlation is accounted for by the correlation between
  the duality properties of $SO(6)$ decuplets and $SO(4)$ triplets within the
  SO(10) self-dual and anti-self-dual
  multiplets  $\Sig,\Sigb$.

\item{c)} $[1,1,-2](\bar F_1,\bar F_2, \bar F_3) \oplus
[1,1,2](F_1,F_2,F_3) \hfil\break
 \equiv (\Sigb_{44(R0)},\phi^{(15)}_{(R-)},\lambda_{(R-)})
\oplus (\s^{44}_{(R0)},\phi^{(15)}_{(R+)},\lambda_{(R+)})$

The mixing matrix $\cal F$ has the usual structure . The residual
massive eigenstates after separating off the the two Dirac
fermions of mass

\be m_{\lambda_F}= g(24|\omega|^2 + 2|\sigma|^2)^{1\over 2} \ee is
a Dirac fermion of mass

\be m_F=|{\eta\over{\omega}}| {\sqrt{ |\sigma|^2 + 12|\omega|^2}}
\ee and the form of its chiral parts is

\bea F&=& N_F (i \sqrt{12} \omega  F_1 +\sigma F_2)
e^{i(\gamma_{\eta}-\gamma_{\omega})}\nonumber \\
{\bar F} &=& N_F
 (-i \sqrt{12} \omega {\bar F}_1 +{\bar \sigma} {\bar F}_2)
  \nonumber\\
N_F^{-1} &=& \sqrt{12|\om|^2 + |\sss|^2} \eea

\item{d)} $[\bar 3,1,-{4\over 3}] (\bar J_1,\bar J_2,\bar J_3,\bar
J_4) \oplus [3,1,{4\over 3}](J_1,J_2,J_3,J_4)  \hfil\break \equiv
(\s^{\bm4}_{(R-)},\phi_4^{~\bm},
\phi_4^{~\bm(R0)},\lambda_4^{~\bm}) \oplus
(\Sigb_{\bm4(R+)},\phi_{\bm}^{~4},\phi_{\bm(R0)}^{~4},
\lambda_{\bm}^{~4}) \qquad ;  $

The $4\times 4$ mass matrix $\cal J$
  has the usual super-Higgs structure : complex conjugates of
 the 4th row and column are left and right null eigenvectors
of the upper left $3 \times 3 $ submatrix ${\bf J}$.
 $\cal J$ has non zero determinant
although the determinant of ${\bf J}$ vanishes.

\item{e)} $ [3,2,{{5}\over 3}]({\ovl X}_1,{\ovl X}_2,{\ovl X}_3)
\oplus    [3,2,-{{5}\over 3}](X_1,X_2,X_3)$\hfil\break $\equiv
(\phi^{\bm4 (s)}_{\alpha\dot 1}, \phi^{\bm4(a)}_{\alpha\dot
1},\lambda^{\bm4}_{\alpha\dot 1}) \oplus (\phi_{\bm4\alpha\dot
2}^{(s)}, \phi_{\bm4\alpha\dot 2}^{(a)},\lambda_{\bm4\alpha\dot
2})$

mix via a $3\times 3 $ symmetric matrix ${\cal X}$
  so the  left and right null eigenvectors of the
 upper left $2 \times 2$ submatrix ${\bf{X}}$,
   formed by the
complex conjugates of the  third row and column of ${\cal X}$, are
the same. Separating off the two Dirac $[3,2,\pm{5\over 3}]$ gauge
fermions of mass

$$ m_{\lambda_X}= g \sqrt{4 |a + \omega|^2 + 2 |p+\omega|^2} $$

one is left with a Dirac fermion of mass

$$ m_X= 2 (2|m +\lambda (a+\omega)|  + |m+\lambda \omega|)
={{2 |m| (2 |x|^2 +|1-x|^2)}\over {|1-x|}}$$

whose chiral parts are also neatly expressed in terms of $x$

\be (X,\ovl X) = {1\over {\sqrt{2|x|^2 +|1-x|^2} }} (
e^{i(\gamma_m-\gamma_{1-x})}(\sq x X_1 + (1-x) X_2) , (\sq x \ovl
X_1 + (1-x) \ovl X_2) ) \ee

\end{itemize}

This concludes our description of the superheavy mass spectrum of
the  minimal susy GUT. As mentioned earlier our results were
calculated in collaboration with the authors of\cite{bmsv} and are
in agreement with  the chiral spectra calculated in \cite{bmsv} :
whose results also confirm our earlier results \cite{alaps} on the
phenomenologically important matrices $\cal H,\cal T $. Moreover
we have evaluated  the mixing of the gauginos with the chiral
fermions explicitly and calculated the gauge spectra and
eigenstates besides furnishing all the couplings in the
superpotential sector explictly. The gauge couplings and the gauge
Yukawa couplings to matter will be be given in the Section V. The
gaugino mixing with the chiral fields
 will be useful to us  when we examine
$B+L$ violation mediated by gaugino exchange  as well by Higgsino
triplet exchange and when one wishes to examine the flow of gauge
couplings past the gauge thresholds.

\section{ RG Analysis}

The first phenomenological success
 of   GUTs  was  the
 1-loop calculation of the
numerical value of Weinberg angle \cite{geoquinwein}.
 This was followed by the
 prediction\cite{marsenj} and then the
verification\cite{amaldi} of  an amazingly exact compatibility
between  UV gauge coupling convergence in the MSSM and the
precision LEP data.
 The   large mass at which the top quark was
 eventually discovered and the
associated large value $Sin^2\theta_W \sim .23$
 verified the originally somewhat far fetched  conjecture
of \cite{marsenj} :  a historical fact that is still not always
 appreciated.
The proposal of Weinberg\cite{weinberg} for calculating threshold
effects within an effective field theory picture using mass
independent renormalization schemes such as the standard
$\overline{MS}$ renormalization scheme was taken up and developed
in detail in \cite{hall}. Thereafter, using these results,
 it was argued\cite{dixitsher}
 that high-precision calculations in
SO(10), and particularly in supersymmetric SO(10) models which
used large representations such as {\bf{210,54,126}} etc,
 {\it{were futile}}. This was due to   the
 huge corrections to the one loop
predictions that  they expected in view of
  the large number of
superheavy fields and the expected span in their masses. It should
be remarked however that without an explicit calculation
cancellations that might naturally
occur would be overlooked. Such
calculations were never done.  These  negative  expectations were
a motivation for  the development\cite{nrenso10} of a whole
 genre of SO(10) models that
 eschewed large representations (and
 thus parameter counting minimality)
 in favour of models with a  plethora of small  representations
  and non-renormalizable  interactions.

 The other approach \cite{aulmoh,ckn,lee,leem,abmrs,abmsv} approach
  has all along been to retain
 renormalizability of the fundamental theory. We regard
 retention of  Higgs multiplets just
 adequate to account for the gauge and
 fermion spectrum via renormalizable couplings as a
 {\it{sine qua non}} for even being clear as
 to what is testable about a given  model.
  The inverse approach where representations
  (``hypotheses'') are multiplied without necessity
   seems  regressive to us.
Thus the proposal of the Susy SO(10) GUT based on the
${\bf{210-126-{\overline{126}}-10}}$\cite{aulmoh,ckn}
 Higgs system as being {\it{the}}  Minimal Susy
 GUT \cite{lee,abmsv}
 must live or die by the criterion :
 {\it{ Are the one loop values of $Sin^2\theta_W $ and
 $M_X$ generically stable against superheavy threshold
calculations ?}}. By ``generically'' we mean : for a non-singular
subset of the parameter space.
 So far this question could not be answered
definitively since no complete mass spectrum was available  in any
Susy SO(10) model to settle the issue.
 Partly this was due to
the lack of accessible techniques to calculate mass spectra and
couplings in these models due to the difficulty in obtaining the
relevant SO(10) ``Clebsches".
 Over the last few years we have
developed\cite{alaps} a complete technology for translating SO(10)
tensor and spinor  labels into those of the unitary labels of the
Pati-Salam maximal subgroup  \PS  of SO(10). This allowed us to
compute first the mass matrices of SM type doublets and proton
decay mediating triplets and then the complete spectrum and
couplings reported in this paper. The partial technology of
\cite{heme,lee} has also been used to compute\cite{bmsv,fm} this
spectrum (but not the couplings). With the correct spectrum in
hand we can apply the standard formulae of Hall \cite{hall} to
compute the changes in the 1-loop GUT predictions as functions of
the few MSGUT parameters ($\xi=M\lambda/m\eta,
\lambda,\eta,\gamma, {\bar\gamma},m,g, M_H $)   which are relevant
at the GUT scale. Thereafter we can  scan
 the parameter space to see how the corrections vary with these parameters.

 A few remarks on the role of the parameters
 are in order. The parameter $\xi= \lambda M/ \eta m$ is the only
  numerical parameter that
 enters into the cubic equation eqn.(\ref{cubic})
 that determines the parameter $x$ in terms of which all the
 superheavy vevs are given. It is thus
 the most crucial  determinant of the mass spectrum.
 The dependence of the threshold corrections on the parameters
 ${\lambda,\eta,\gamma,{\bar\gamma}}$ seems quite mild (logarithmic)
  (this is especially obvious  for the unmixed chiral multiplets)
 and is also suggested by our preliminary scans of the parameter
 space. Thus changing $\lambda,\eta$ by a factor of 100 each
 yields plots vs $\xi$ that seem indistinguishable from the ones
 presented below.
From equations (\ref{dlvevs}) we see that
  $m/\lambda$ can be extracted as the overall scale parameter
  of the vevs.   Since the threshold corrections
  we calculate are dependent  only on (logarithms of)
  ratios of masses
  the parameter $m$ does not play any crucial role in our
  scan of the parameter space : it is simply fixed
  in terms of the (threshold and two loop corrected)
  mass $M_V=M_X$
  of the lightest superheavy vector particles mediating
  proton decay : which mass is chosen, in the approach of Hall,
   as the common
  ``physical" matching point in the equations relating
  the running MSSM couplings to the SO(10)
  coupling\cite{hall}. Inasmuch as we take the
 parameters $\lambda,\eta$ as given,
  and the parameter $m$ is set by the overall mass scale,
  the freedom in the parameter $\xi$
 is essentially that  of choosing the ${\bf{126-\oot}}$
 mass  parameter
 $M$ i.e the freedom in choosing the dimensionless parameter $\xi$,
  is essentially that of the
 ratio $M/m$ : which ratio is already known to be a
 crucial control parameter of symmetry breaking in renormalizable
 models that utilize the ${\bf{126, \oot}} $
to complete and enforce the symmetry breaking down
 to the SM symmetry\cite{abmrs}. As
for $M_H$ it is fine tuned to keep a pair of doublets light. The
relation between the MSSM couplings at the susy breaking scale
$M_S\sim 1 $TeV and the GUT coupling at the
    scale $M_X$ is given by \cite{weinberg,hall} :

\bea {1\over{\alpha_i(M_S)}}
 ={1\over{\alpha_G(M_X)}} +
 8 \pi b_i ln{{M_X}\over{M_S} }
  + 4 \pi \sum_j {{{b_{ij}} \over {b_j}}} ln X_j -
4\pi\lambda_i(M_X)
\eea

here

 \be X_j= 1 + 8 \pi b_j \alpha_G(M_X^0)
 ln{{M_X^0}\over{M_S }}
\ee

is understood to be evaluated at the values of
$M_X^0,\alpha_G(M_X^0)$ determined from the one loop calculations.
In this equation the contribution of the yukawa couplings has not
been taken into account and this should also be done in a full
investigation \cite{abmsv2}. Here we will confine ourselves to
estimating the corrections using the equations as given above ,
since these were already conjectured\cite{dixitsher}  to lead to a
breakdown of the unification scenario. The coefficients

\be \{b_1,b_2,b_3\}=(1/16\pi^2) \{{{33}\over 5},1,-3\}\nonumber
\ee \be
 [b_{ij}]={1\over{(16\pi^2)^2}}
  \left({\begin{array}{ccc}{{199}/{25}}  &
{{27}/{5}}  & {{88}/{5}}\\
 {9/ 5} &25 & 24 \\
 {{ 11}/{5}} & 9& 14
 \end{array}}\right)
\ee

  are the standard one loop and two loop gauge
evolution coefficients for the MSSM \cite{martin}. The term
containing $\lambda_i$ represents the leading contribution of the
superheavy thresholds :

\be \lambda_i (\mu)=-{2\over {21}} (b_{iV} + b_{iGB})
 + 2(b_{iV} + b_{iGB})ln{{M_V}\over{\mu }} +2 b_{iS}ln{{M_V}\over{\mu }}+2
b_{iF}ln{{M_F}\over{\mu }} \ee

where V,GB,S,F refer to vectors, Goldstone bosons,
 scalars and fermions respectively and a sum over heavy
 mass eigenstates is implicit. The formulae for the threshold corrections are

\bea \Delta^{(th)}(ln{M_X}) &=&{{5\lambda_1(M_X^0) +
3\lambda_2(M_X^0) - 8\lambda_3(M_X^0)} \over{10b_1+6b_2-16b_3}}
  \\
 \Delta^{(th)}(Log_{10}{M_X})  &=& .0217 +.0167 (5 {{\bar b}'}_1 +3{{\bar b}'}_2 -8
  {{\bar b}'}_3) Log_{10}{{M'}\over  {M_X^0}} \label{Deltasw}\\
\Delta^{(th)} (sin^2\theta_W (M_S)) &=&
 {{10\pi\alpha(M_S)}\over{(5b_1+3b_2-8b_3)}}
 \sum_{ijk}\epsilon_{ijk}(b_i-b_j)\lambda_k(M_X^0)\nonumber\\
&=& .00004 -.00024 (4 {{\bar b}'}_1 -9.6 {{\bar b}'}_2 +5.6
  {{\bar b}'}_3) Log_{10}{{M'}\over  {M_X^0}}
  \label{Deltath} \eea
Where ${\bar b'}_i =16 \pi^2 b_i'$ are the 1-loop beta function
coefficients for multiplets with mass $M'$. To evaluate these
formulae it is convenient to group the gaugino contributions along
with the chiral fermions they  mix with. The values of the indices
$S_1,S_2,S_3$ combined as in eqns.(\ref{Deltasw}, \ref{Deltath})
i.e $ S_W=4 S_1-9.6 S_2 +5.6 S_3 ; \quad  S_X= 5 S_1+3 S_2-8 S_3 $
are given in  Table 2 in the appendix.

 The two loop contributions

 \bea \Delta^{(2-loop)}(ln{M_X}) &=&
 -{1\over{10b_1+6b_2-16b_3} } \sum_j[{{5b_{1j}+ 3 b_{2j}-8b_{3j}}  \over{b_j}}
 ln X_j]\nonumber\\
\Delta^{(2-loop)} (sin^2\theta_w (M_S)) &=&
 -{{10\pi\alpha(M_S)}\over{(5b_1+3b_2-8b_3)}}
\sum_{ijkl}\epsilon_{ijk}(b_i-b_j)
 {{b_{kl}}\over {b_l}}ln X_l
\eea
  Using the values
  \bea
    \alpha_G^0(M_X)^{-1}= 25.6\quad ; \quad M_X^0=
10^{16.25} GeV \quad ;\quad
 M_S=1 TeV \nonumber \\
 \alpha_1^{-1}(M_S)=57.45 \quad ;\quad  \alpha_2^{-1}(M_S)=30.8 \quad ;
 ; \quad\alpha_3^{-1}(M_S)=11.04
 \eea
 extrapolated from the global averages
of current data, the two loops effects give

\be
 \Delta^{2-loop}(log_{10}{{M_X}\over {M_S}})=-.08 \qquad ;
  \quad \Delta^{2-loop}(sin^2\theta_W (M_S))= .0026
  \ee

The values of the 1-loop coefficients
${\bar b_i}=16\pi^2b_i$
corresponding to Vector, complex scalar and Weyl fermion fields
are $-11S(R)/3, S(R)/3, 2S(R)/3 $ where $S(R)$ is the index of the
relevant representation. Note in particular that this implies that
the nonzero hypercharge superheavy vector multiplets   which are
present in SO(10) models will contribute with {\it{negative}}
coefficients to the evolution of even the $U(1)_Y$ coupling.

 We have computed the threshold corrections
 for a range of values of $\xi$ keeping the other ``insensitive"
 parameters   fixed at  randomly chosen
 representative values

 \be \lambda =0.12 \quad ; \quad \eta =0.21\quad ; \quad\gamma=0.23\quad ; \quad
 \bar\gamma=0.35 \ee

 The results for different values of these parameters (but with the same
 $\xi$)   are very similar.
 We will therefore keep them fixed at these
 values throughout since here we only wish to illustrate the
 feasibility of precision RG calculations in the SO(10) MSGUT.

  For real values of the superpotential
 parameters the  cubic equation  (\ref{cubic})
 that determines the vevs has one real and two complex (conjugate)
solutions. The latter give essentially identical corrections. So
for real $\xi $ we need to present plots for two solutions only.
These are given as Figs. 1-6.

\begin{figure}[h!]
\begin{center}
\epsfxsize15cm\epsffile{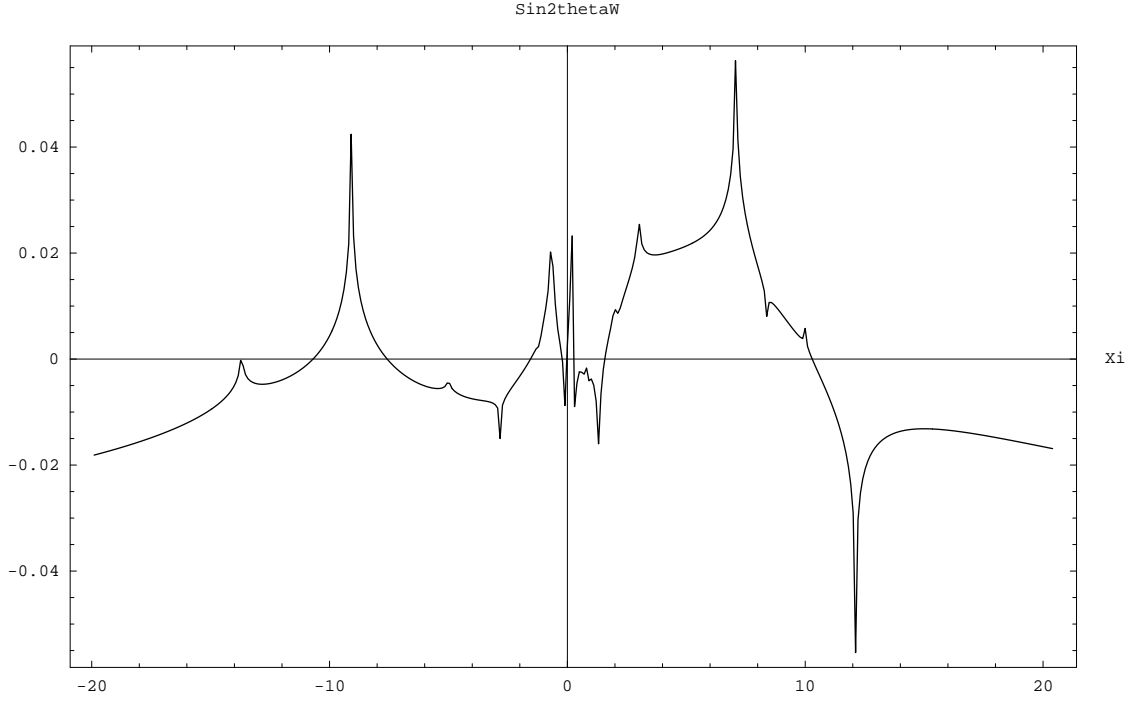}
 \caption{  Plot of the threshold
corrections to $Sin^2\theta_w$ vs $\xi$ for real $\xi$ : real
solution for x.}
\end{center}
\end{figure}

\begin{figure}[h!]
\begin{center}
\epsfxsize15cm\epsffile{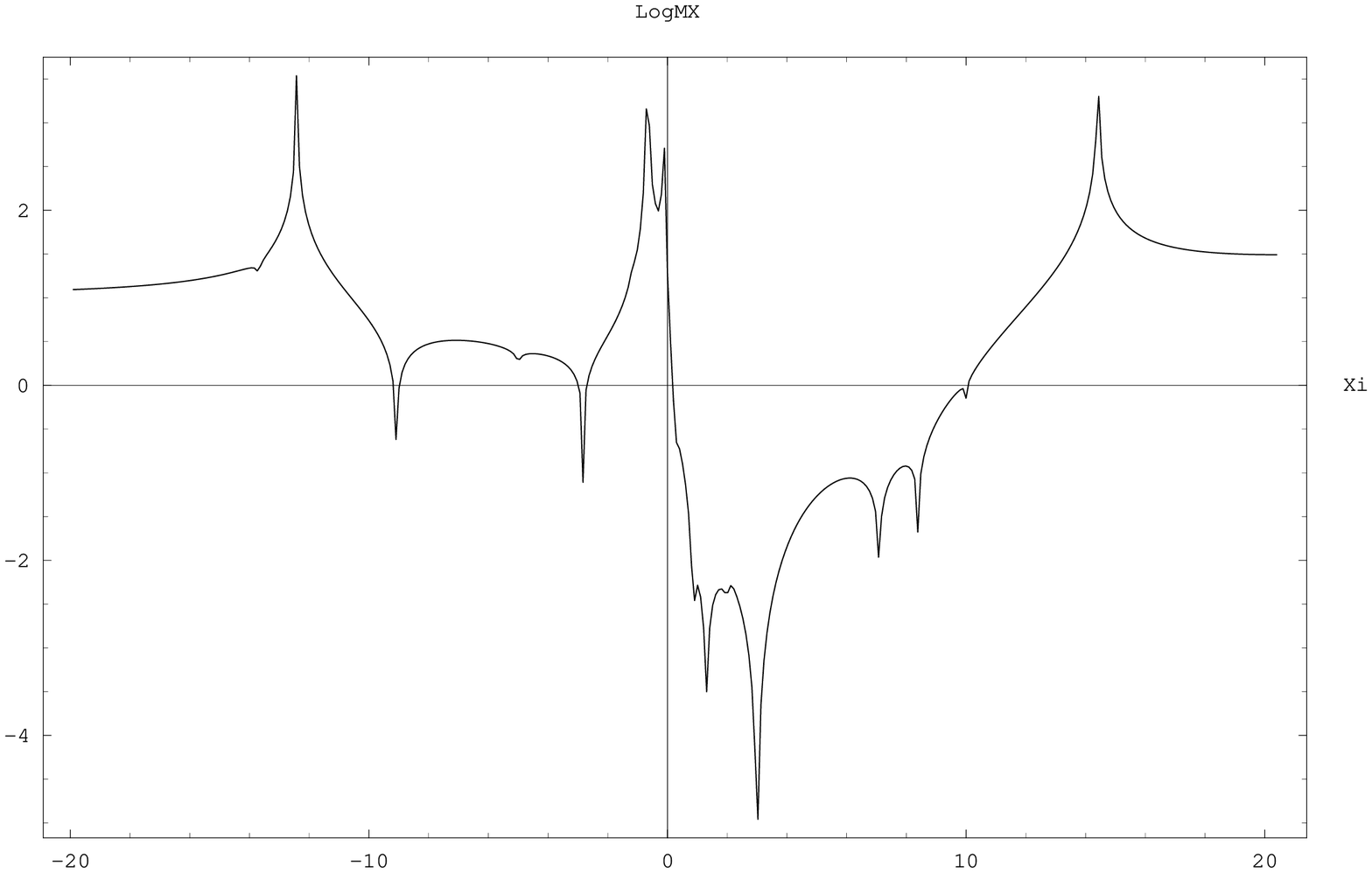}
 \caption{  Plot of the threshold
corrected   $Log_{10} M_X/M_X^0 $ vs $\xi$ for real $\xi$ : real
solution for x.}
\end{center}
\end{figure}

From  Figs. 1,3  we see that for most real values of $\xi $ the
threshold effects on $sin^2\theta_W (M_S)$ are less than 10 \% of
the 1-loop values. There are three exceptional values of $\xi $
very close to which
 this limit is breached but even then the change is only about
  25\%.
For large magnitudes of $\xi$ an asymptotic regime of around 10\%
change seems to  supervene. Similarly Figs. 2,4
 show  that the change in $M_X$ is also
 not drastic(though possibly
  phenomenologically interesting
 since the gauge contribution to the
 nucleon lifetime goes as $M_X^{-4}$)
 except at certain
 special  points among which one recognizes certain known points
 of enhanced symmetry\cite{abmsv,bmsv} such as
   $\xi=-5,10$ (SU(5)), $\xi=3$ ($G_{LR}$), $\xi=-2/3$
    (flipped $SU(5)\times U(1)$).
     It is natural to expect that
    something similar accounts for the other sharp peaks and dips
    in these plots. Moreover their narrowness emphasizes that for
    generic values of the parameters one may expect the threshold
    corrections to be small for the real $\xi$ real  $x$
    cases.  There are also
     regions in which the threshold corrections to
     $Log_{10}M_X$ are as large as $-5$ and
     these need special examination with regard
     to their phenomenological viability and
     consistency with the one scale breaking picture.
      It is interesting that in this way
    one can scan the parameter space of the MSGUT and obtain
    a global``tomograph"  of the variation in its  character
    with the ratio $M/m$.

\begin{figure}[h!]
\begin{center}
\epsfxsize15cm\epsffile{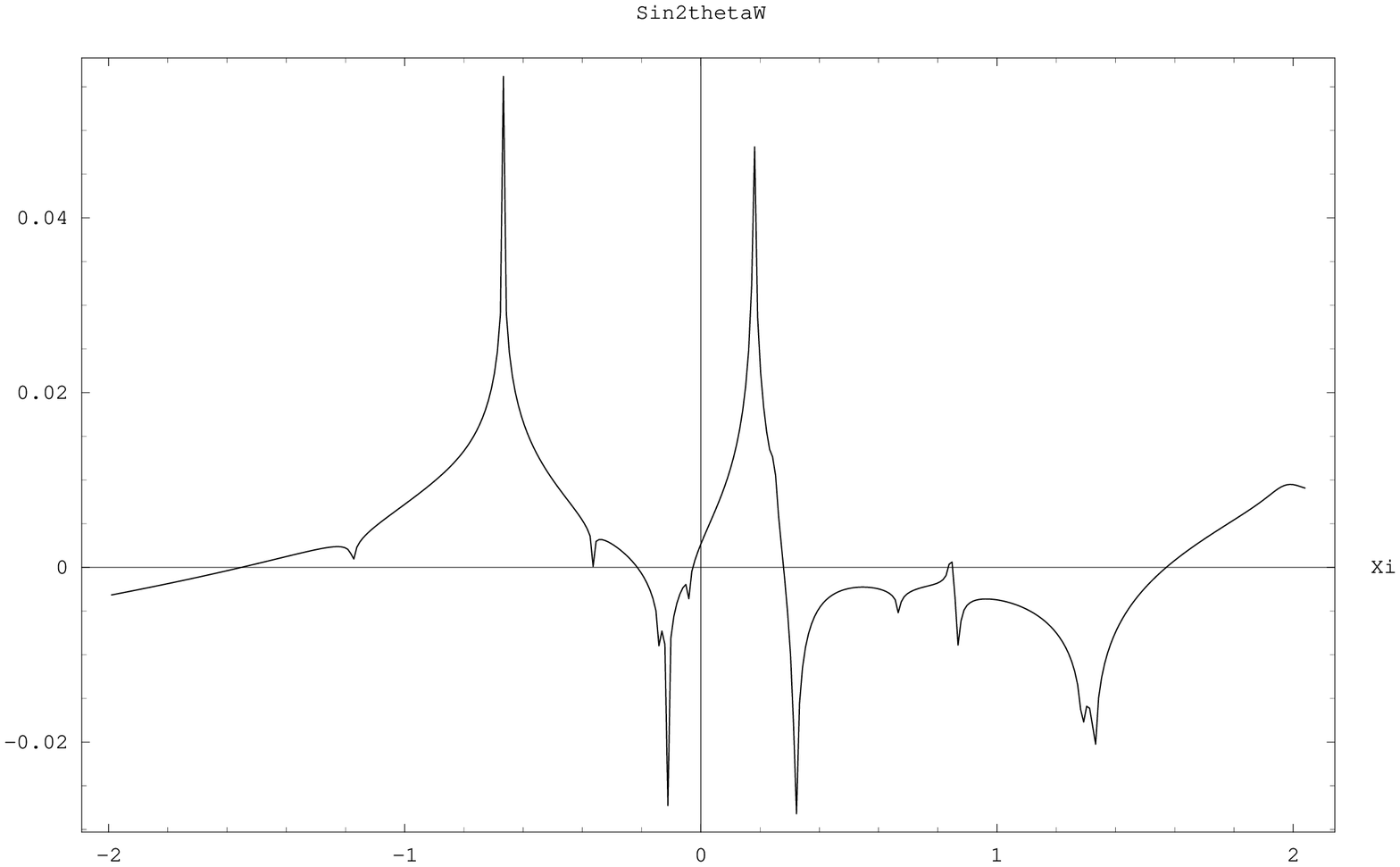}
 \caption{  Magnified Plot of the threshold
corrections to $Sin^2\theta_w$ vs $\xi$ for
 real $\xi$ : real
solution for x.}
\end{center}
\end{figure}

\begin{figure}[h!]
\begin{center}
\epsfxsize15cm\epsffile{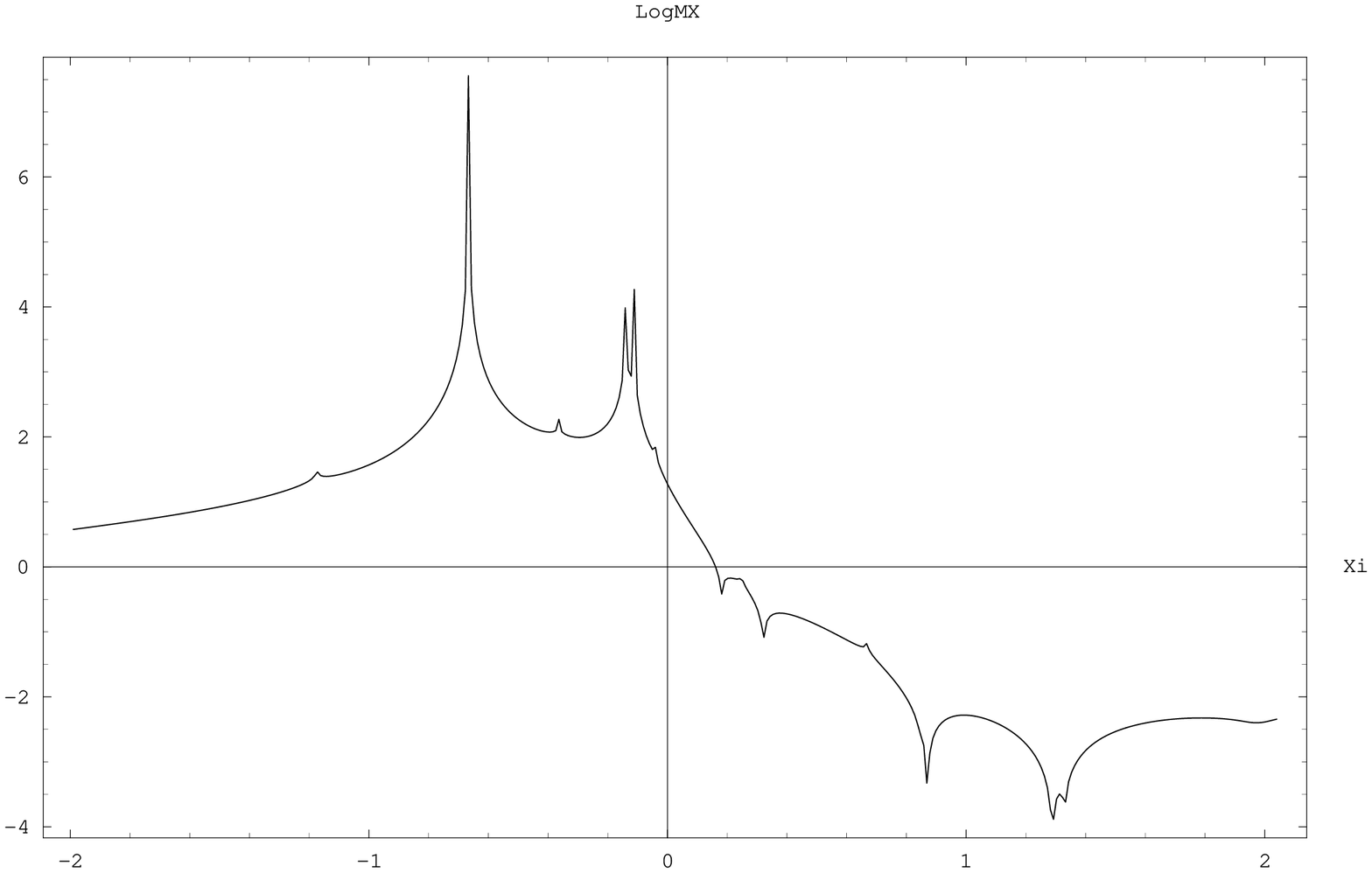}
 \caption{  Magnified Plot of the threshold
corrected   $Log_{10} M_X/M_X^0 $ vs $\xi$ for real $\xi$ : real
solution for x.}
\end{center}
\end{figure}

 Fig. 3., 4. we give a magnified view of
 the region  $  |\xi| < 2 $. A comparison of the graphs shows
 clearly that the peaks in the threshold corrections coincide
 by either measure, obviously because some particles are
 becoming very light and enhancing the mass ratios that enter
 the formulae. It will be amusing to use these plots to
identify and unravel the special regions of the MSGUT parameter
space.

\begin{figure}[h!]
\begin{center}
\epsfxsize15cm\epsffile{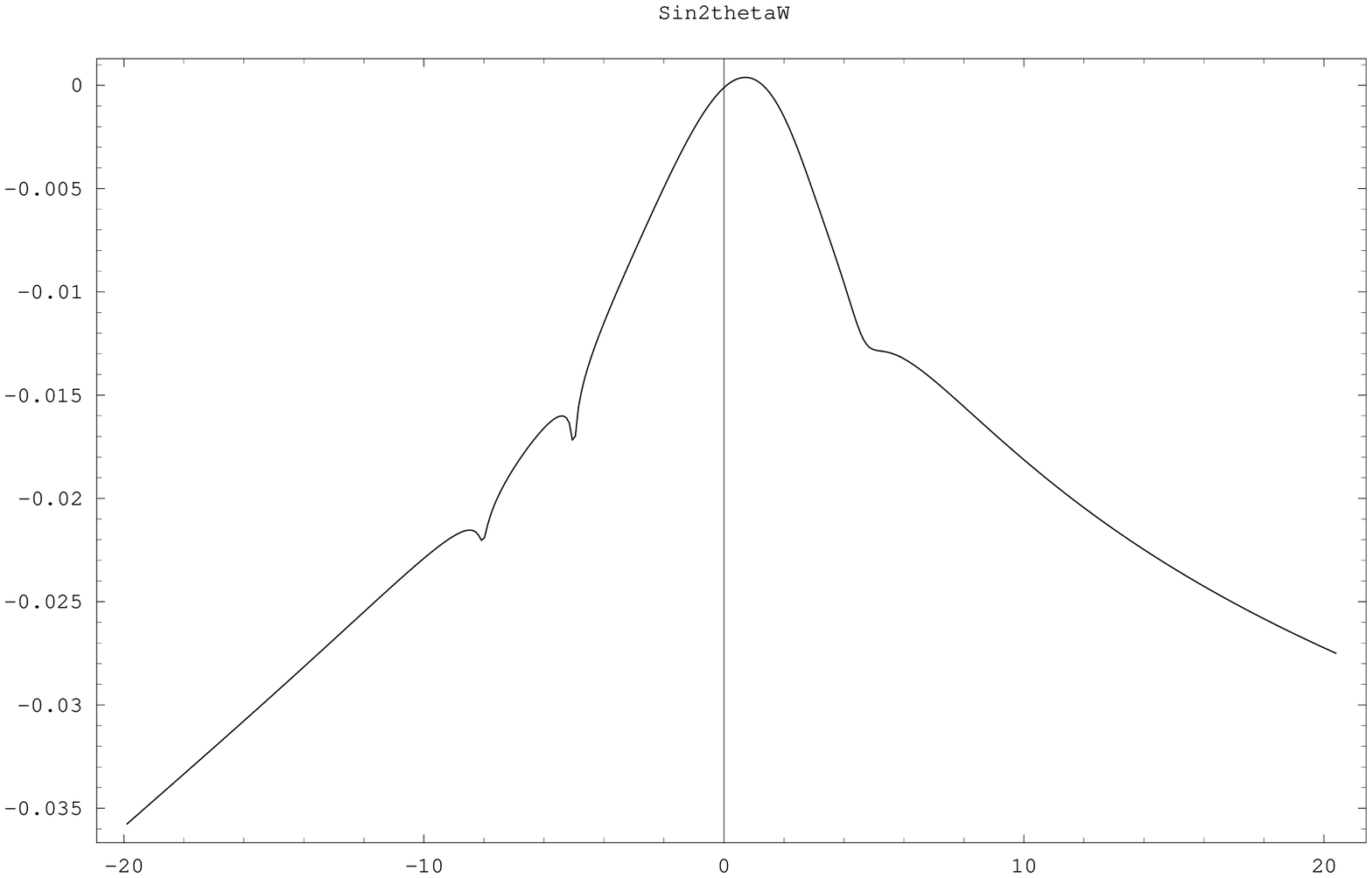}
 \caption{    Plot of the threshold
corrections to $Sin^2\theta_w$ vs $\xi$ for
 real $\xi$ :  complex
solution for x.}
\end{center}
\end{figure}

\begin{figure}[h!]
\begin{center}
\epsfxsize15cm\epsffile{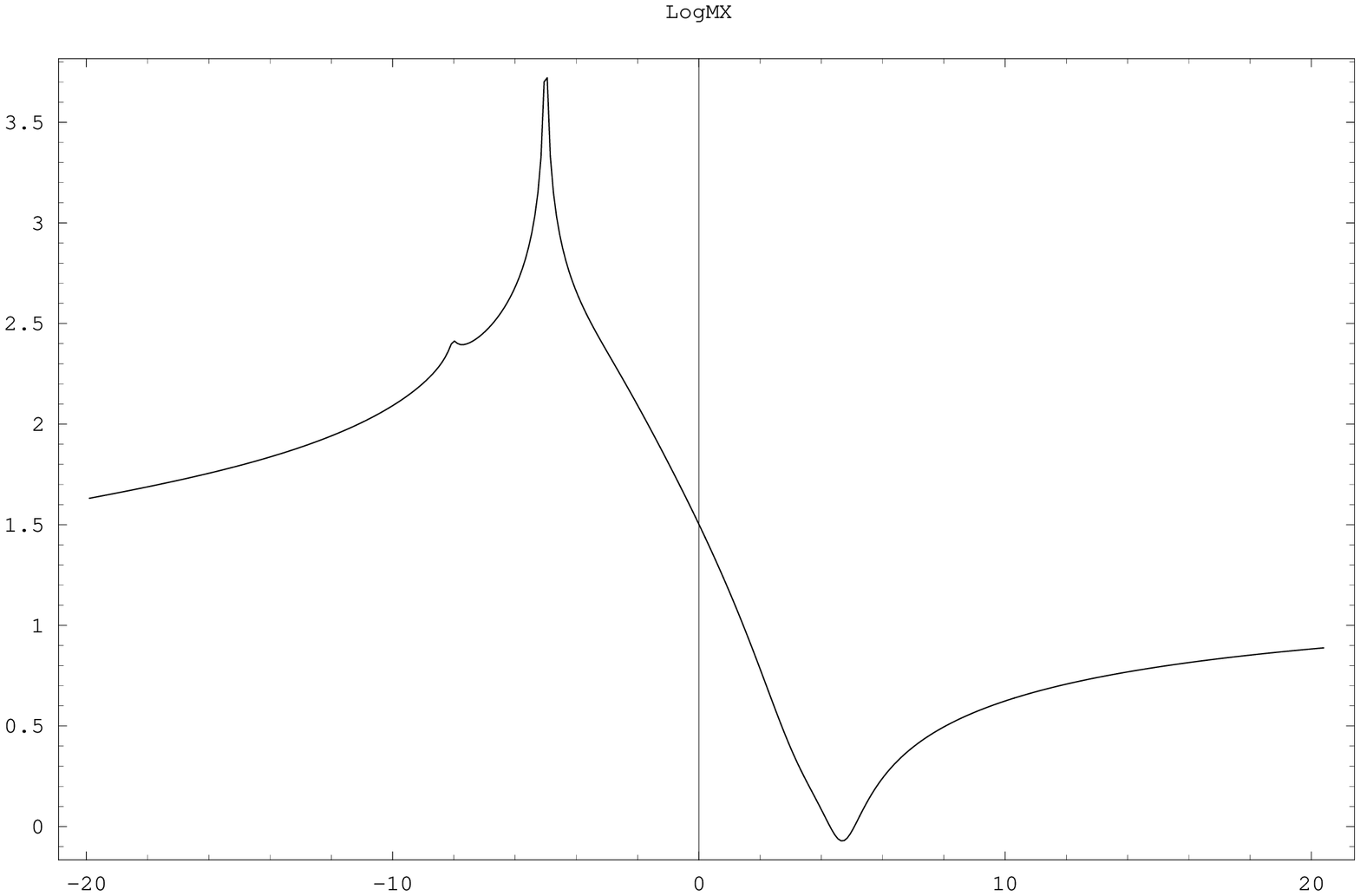}
 \caption{    Plot of the threshold
corrected   $Log_{10} M_X/M_X^0 $ vs $\xi$ for real $\xi$ :
complex solution for x.}
\end{center}
\end{figure}

 Let us turn next to the complex solutions of the cubic
 equation for $x$ but still with real values of $\xi$.
 We obtain the typical plots Fig. 5,6 The
 corrections to $sin^2\theta_W (M_S)$  are very small for small
$|\xi|<2$, with a   minimum close to $\xi =1$. From Fig. 6 we see
that apart from the two peaks near $\xi=\pm 5$ the corrections to
the unification scale are quite small for small $\xi $

\begin{figure}[h!]
\begin{center}
\epsfxsize15cm\epsffile{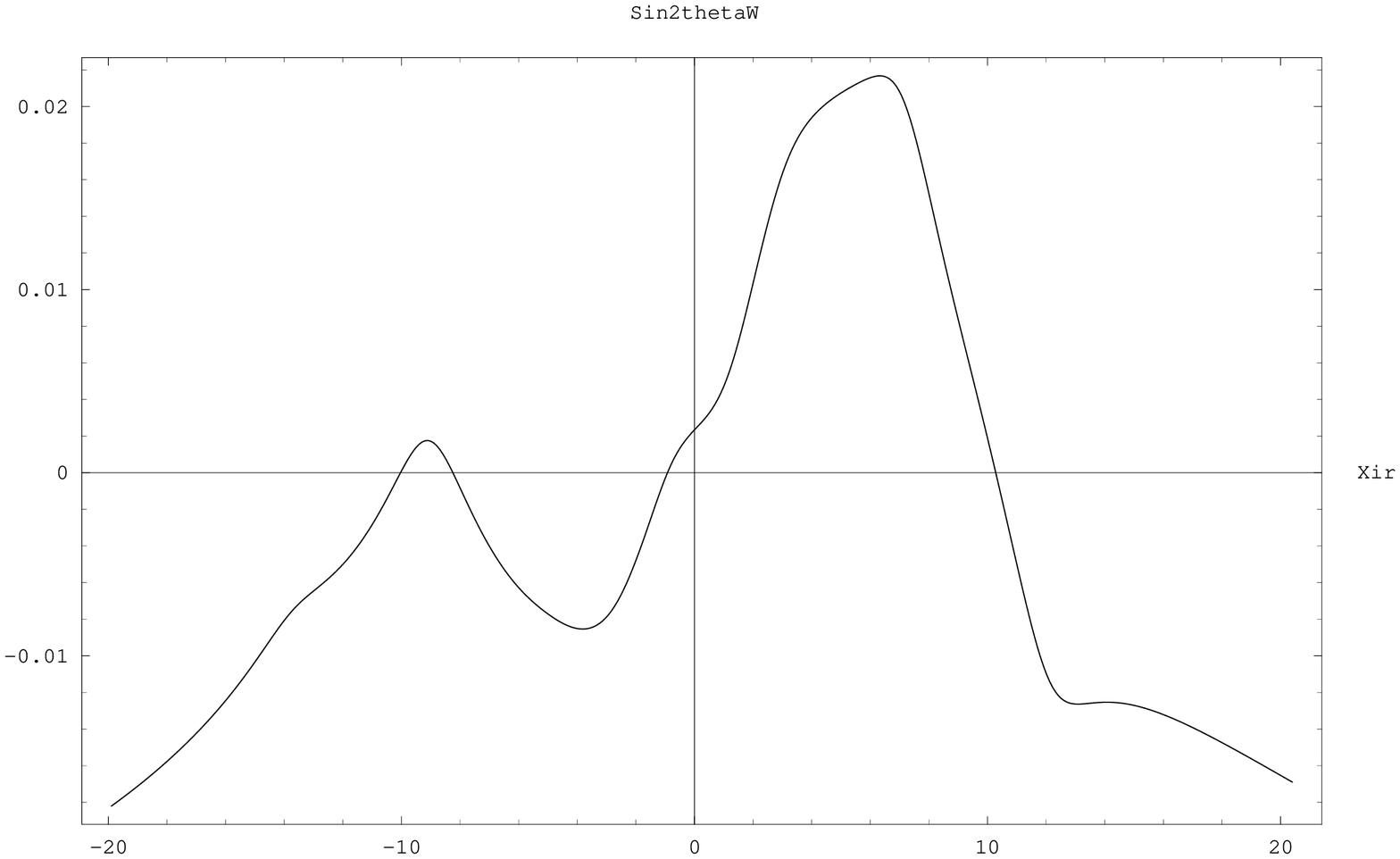}
 \caption{    Plot of the threshold
corrections to $Sin^2\theta_w$ vs $Re(\xi)$ for
 complex $\xi$ : $Im \xi =1.2 $ , first
solution for x.}
\end{center}
\end{figure}

\begin{figure}[h!]
\begin{center}
\epsfxsize15cm\epsffile{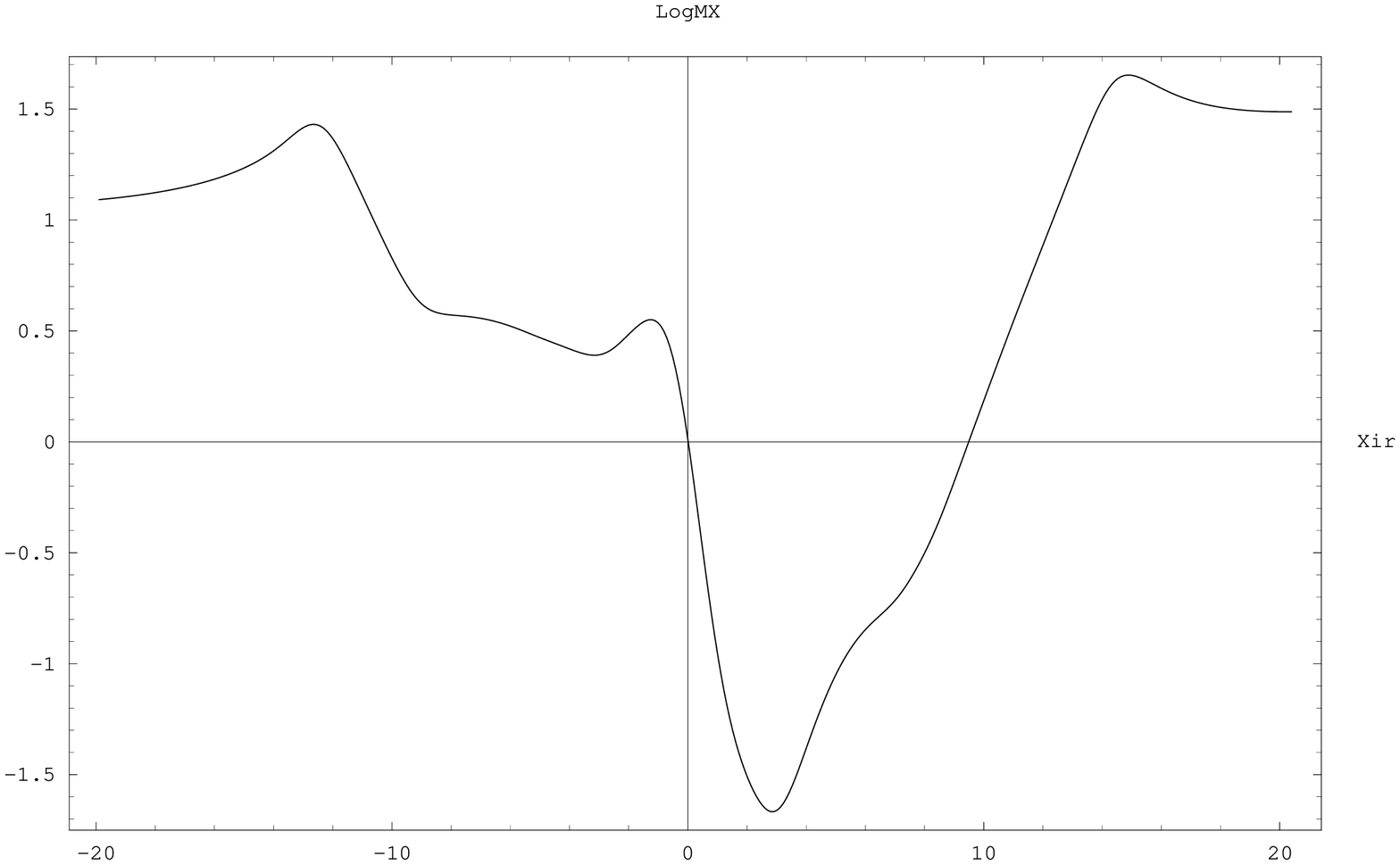}
 \caption{    Plot of the threshold
corrected   $Log_{10} M_X/M_X^0 $ vs  $Re(\xi)$ for
 complex $\xi$ : $Im \xi =1.2 $ , first
solution for x.}
\end{center}
\end{figure}

When we consider complex values of $\xi$ as shown in Fig.7-12 we
see that the behaviour is quite regular (like the case of the
complex solutions for real $\xi $) and once again there are large
regions of parameter space where the corrections are less than
10\% for $Sin^2\theta_w$ while $M_X$ changes by a factor of 10 or
less. Thus even this cursory scan of the MSGUT parameter space
shows that, quite contrary to expectations  in the
literature\cite{dixitsher},
 precision RG analysis of the SO(10)
MSGUT  is  far from being futile,  since the hierarchy of
magnitudes between MSSM one loop gauge coupling convergence values
($O(\alpha^{-1})$ effects ) and the one loop threshold  and two
loop gauge coupling corrections
($O(1)$ effects \cite{hall}) is
generically maintained at the level of 10\% or less. Furthermore
the RG analysis and parameter scan in terms of the single
parameter $\xi $ can teach us much about the structure of the
parameter space since it shows a sharp sensitivity to points of
enhanced symmetry. We will return to these questions at length in
the sequel \cite{abmsv2}.

\begin{figure}[h!]
\begin{center}
\epsfxsize15cm\epsffile{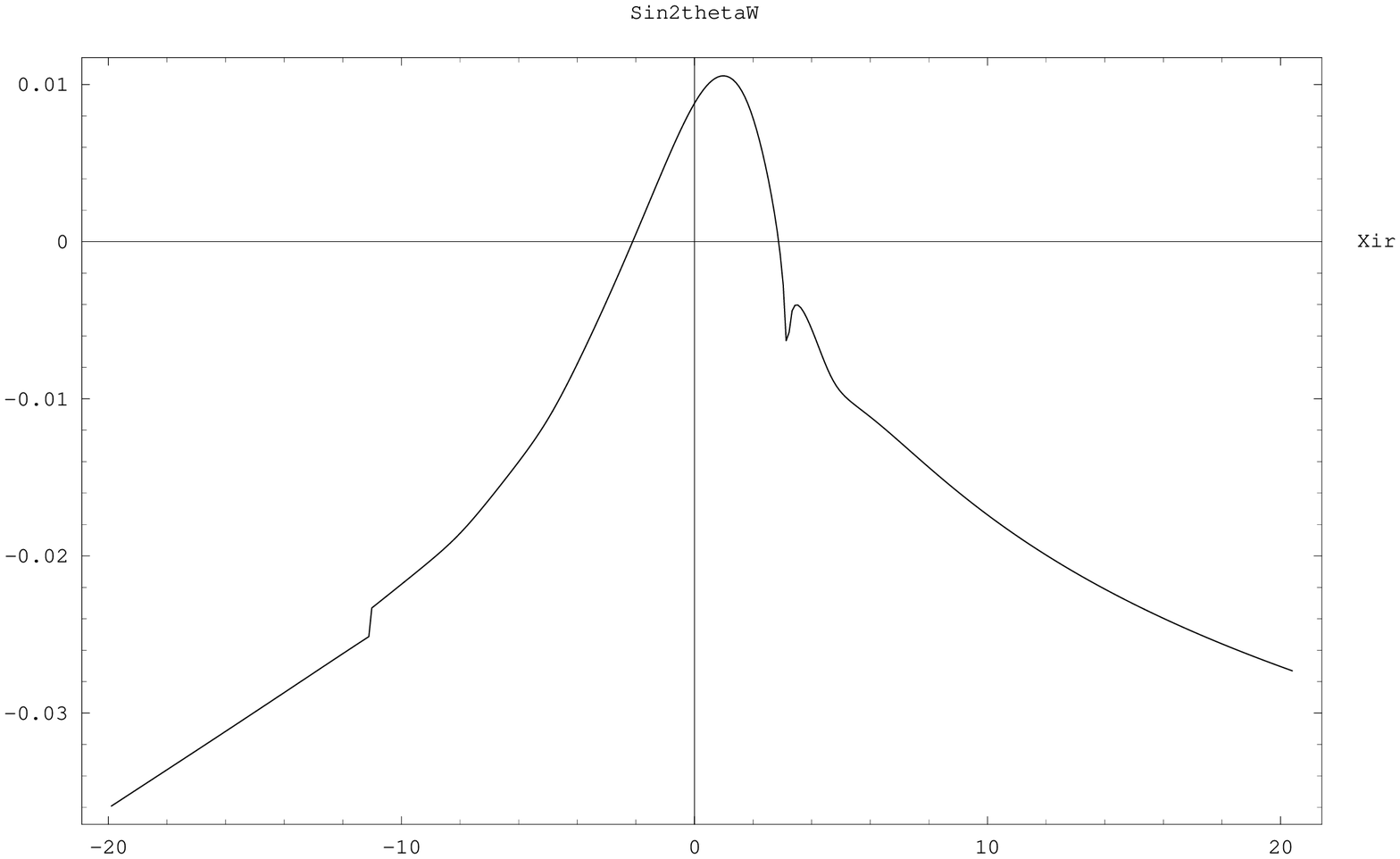}
 \caption{    Plot of the threshold
corrections to $Sin^2\theta_w$ vs $Re(\xi)$ for
 complex $\xi$ : $Im \xi =1.2 $ , Second
solution for x.}
\end{center}
\end{figure}

\begin{figure}[h!]
\begin{center}
\epsfxsize15cm\epsffile{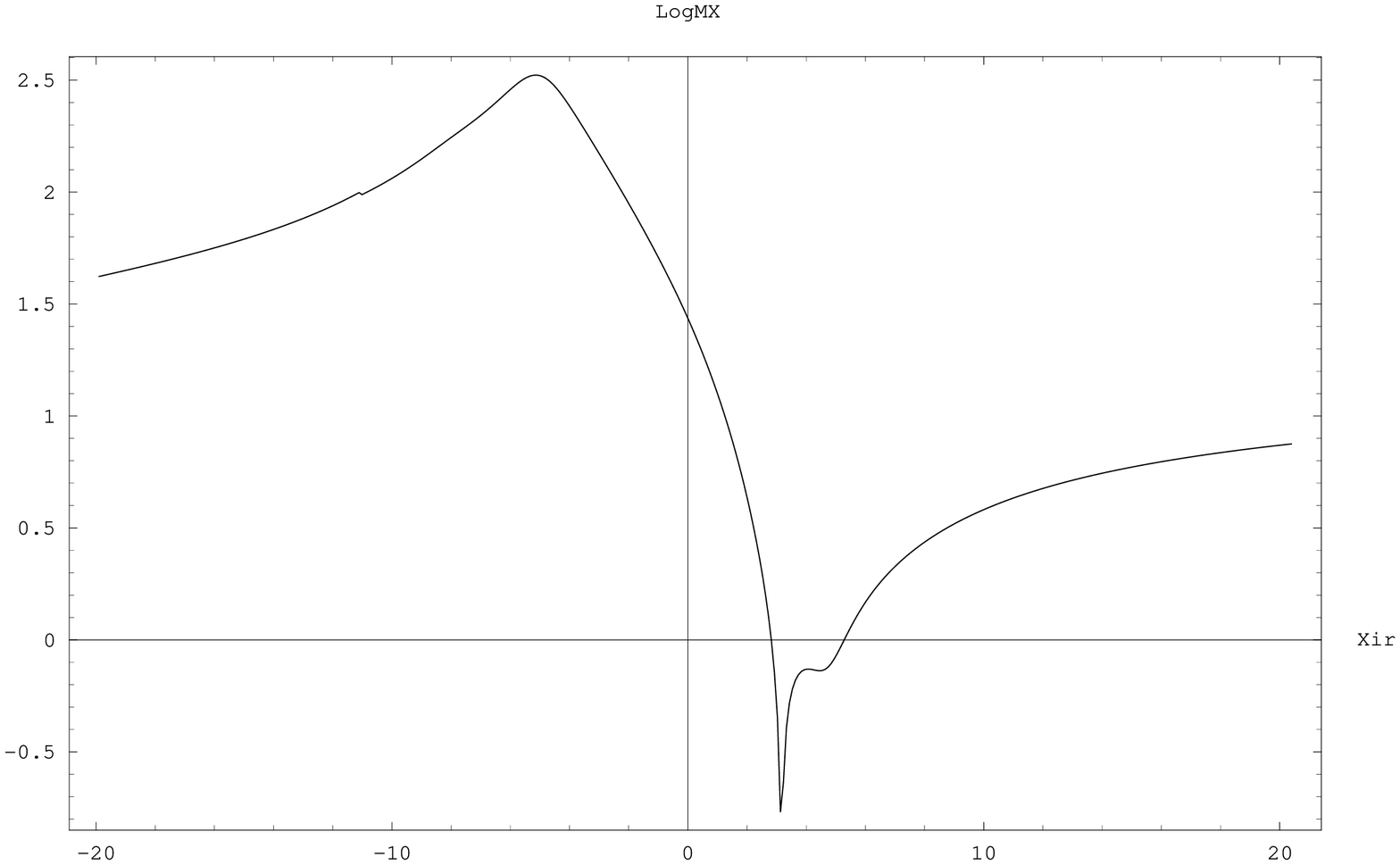}
 \caption{    Plot of the threshold
corrected   $Log_{10} M_X/M_X^0 $ vs $Re(\xi)$ for
 complex $\xi$ : $Im \xi =1.2 $ , Second
solution for x.}
\end{center}
\end{figure}

\begin{figure}[h!]
\begin{center}
\epsfxsize15cm\epsffile{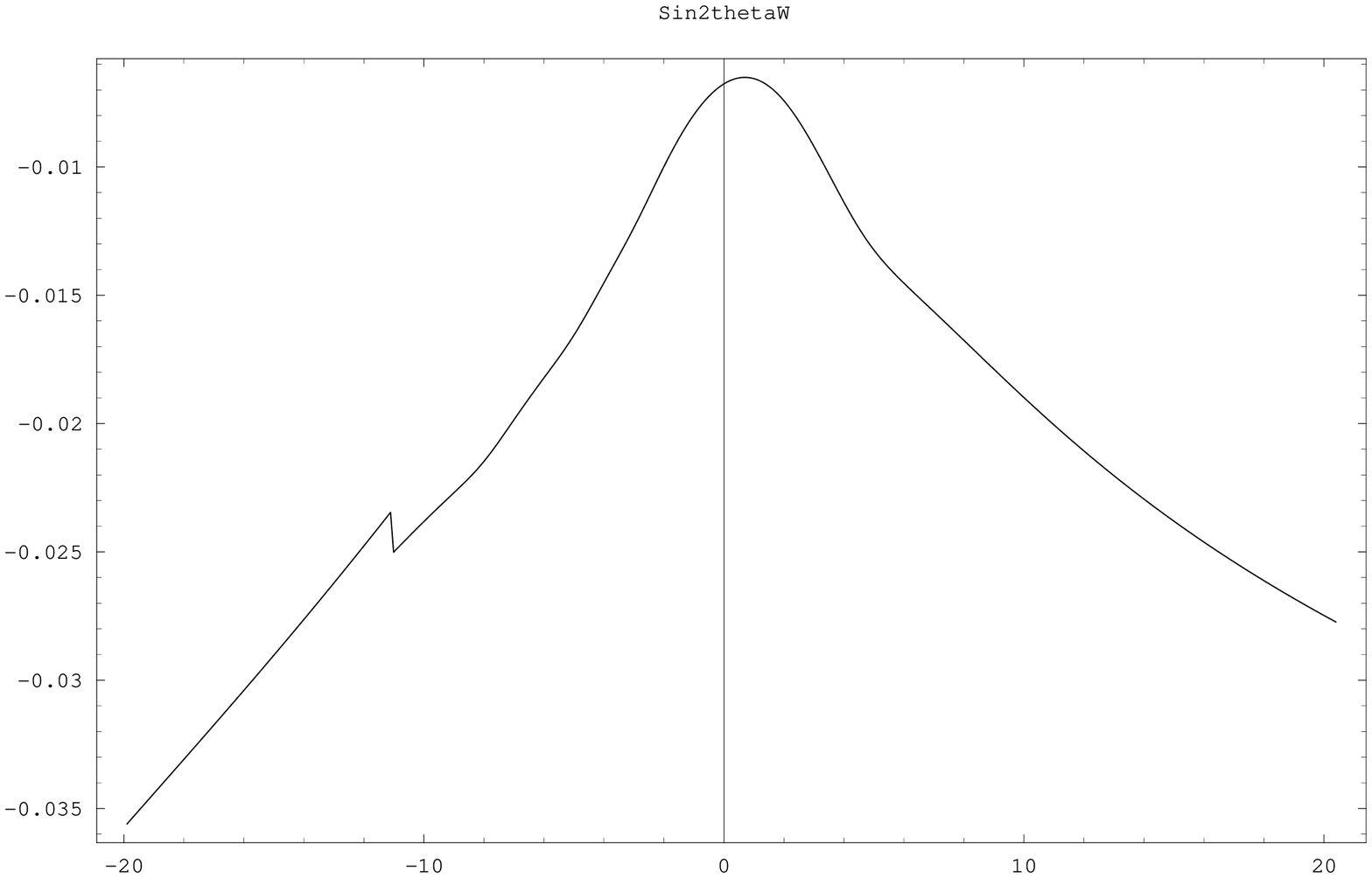}
 \caption{    Plot of the threshold
corrections to $Sin^2\theta_w$ vs $Re(\xi)$ for
 complex $\xi$ : $Im \xi =1.2 $ ,  Third
solution for x.}
\end{center}
\end{figure}

\begin{figure}[h!]
\begin{center}
\epsfxsize15cm\epsffile{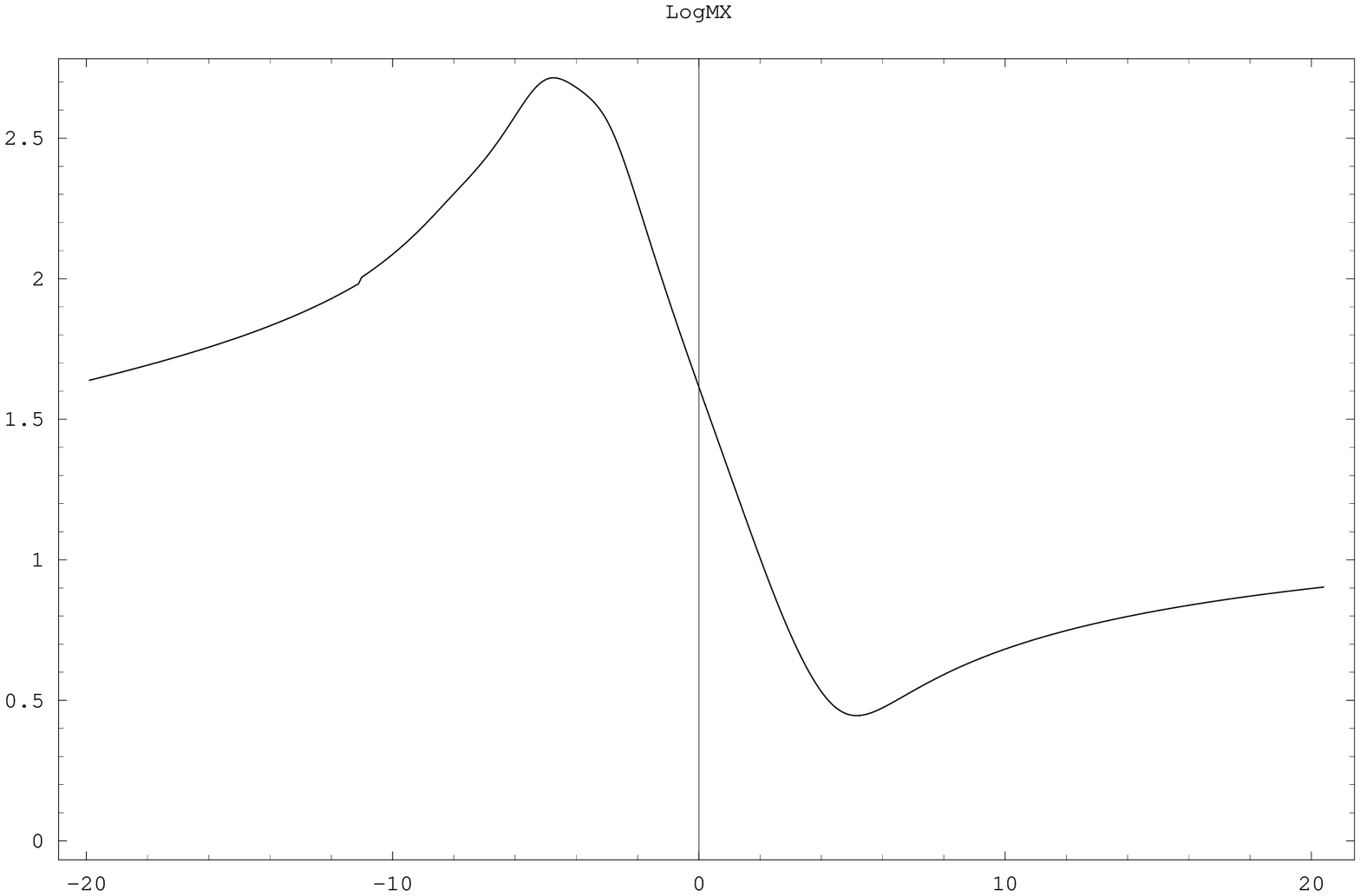}
 \caption{    Plot of the threshold
corrected   $Log_{10} M_X/M_X^0 $ vs $Re(\xi)$ for
 complex $\xi$ : $Im \xi =1.2 $ ,  Third
solution for x.}
\end{center}
\end{figure}

\section{$d=5 $ Operators for B,L violation}
\subsection{Higgsino exchange mediated violation}

 In SO(10) $B-L$ is a gauge symmetry. Thus the vertices preserve
this symmetry and the leading effects of the spontaneous violation
of $B-L$ by the superheavy ${\bf{\s ,\Sigb}}$ vevs are just the
neutrino mass phenomena. To examine the generation of effective
(non-renormalizable) operators that violate $B+L$ in the low
energy theory via Higgsino and gaugino exchange we need the MSSM
break-up of the SO(10) invariants $16\cdot 16\cdot 10$ and
$16\cdot 16 \cdot {\overline{126}}$. Since we had already
presented  the PS decomposition of these invaraiants in
\cite{alaps} it is a trivial exercise to use that result to obtain
the MSSM wise  decompositions :

\bea W_{FM}^H &=& h_{AB}' \psi^{T}_{A}{C}_{2}^{(5)}
\gamma_{i}^{(5)}\psi_{B} H_{i} \nonumber\\
&=& \sqrt{2}h_{AB}' \big [H_{\mu\nu}\widehat{\psi}^{\mu\da}_{A}
\widehat{\psi}^{\nu}_{B\da} +\widetilde{H}^{\mu\nu}\psi_{\mu
A}^{\alpha} \psi_{\nu\alpha B} - H^{\alpha\dot\alpha}
(\widehat{\psi}^{\mu} _{A\dot\alpha}\psi_{\alpha\mu B}
+\psi_{\alpha\mu A}\widehat{\psi}
_{\dot\alpha B}^{\mu} )\big ]\nonumber\\
&=& 2\sqrt{2} h_{AB}'[ \ovl{t}_{1} ({\epsilon} {\bar u}_A {\bar
d}_B+ Q_{A}L_{B})
 + {t}_{1} ( {\epsilon\over 2} Q_{A}Q_{B}+
   \bar u_{A}{\bar e}_{B}-{\bar d}_{A}{\bar {\nu}}_{B})]
\nonumber\\
&-&2\sqrt{2}h_{AB}'\bar{h}_{1}[{\bar d}_{A} Q_{B} +\bar
e_{A}L_{B}] +2\sqrt{2}h_{AB}'{h}_{1}\big [{\bar u}_{A} Q_{B}
+{\bar\nu}_{A}L_B \big] \eea
 %
%

 \bea   W_{FM}^{\Sigb} &=&   {1 \over 5!}
\psi^{T}C_{2}^{(5)}\gamma_{i_{1}}.....\gamma_{i_{5}}\chi
{\overline\Sigma}_{i_{1}...{i_{5}}} \nonumber\\
&=&2\sqrt
{2}{(\widetilde{{\overline\Sigma}}^{\mu\nu}_{(a)}\psi_{\mu}^
{\alpha}\chi_{\nu\alpha}-{\overline\Sigma}_{\mu\nu}^{(a)}\widehat
\psi^{\mu\dot\alpha}{\widehat
 \chi}_{\dot\alpha}^{\nu})}
+{4 \sqrt {2}}{\overline\Sigma}^{~\mu\alpha\dot\alpha}_{\nu}
 {(\widehat\psi_{\dot\alpha}^{\nu}\chi_{\alpha\mu}+\psi_{\mu\alpha}\widehat
\chi_{\dot\alpha}^{\nu})} \nonumber\\
&+&4({\overline\Sigma}_{\mu\nu}^{\dot\alpha\dot\beta}\widehat\psi_{\dot\alpha}^{\mu}
\widehat\chi_{\dot\beta}^{\nu}+{{\overline\Sigma}}^{\mu\nu\alpha\beta}\psi_{\mu\alpha}
\chi_{\nu\beta})\label{ps126yk} \eea

 whence

\bea W_{FM}^{\Sigb}  &=& 4{\sqrt{2}}f_{AB}'
 {  [} t_{2} ({\epsilon
\over 2}Q_{A}Q_B - {\bar u}_{A} {\bar e}_{B}+
 {\bar\nu}_{A} {\bar
d}_{B}) +\bar{t}_{2} ( Q_A L_B - {\epsilon}{\bar u}_{A}{\bar
d}_{B}){  ]}
  \label{WSBFM}\nonumber \\
&+& 4{\sqrt{2}}f_{AB}'
 [{i\over\sqrt{3}} \{ \bar{h}_{2}({\bar
d}_{A}Q_{B}- 3{\bar e}_{A} L_{B})
 -{h}_{2}({\bar u}_{A}Q_{B}
-3{\bar\nu}_{A}L_{B})\}
  \nonumber\\
&+&2 (\bar C_1\bar{d}_{A} Q_{B}-C_2\bar{u}_{A}Q_{B}) +2
(E_{1}\bar{d}_{A}L_{B}-D_2\bar{u}_{A}L_{B})
\nonumber\\
&+&  2(\bar D_{2} {\bar e}_{A}Q_{B}-
\bar E_{2} \bar\nu_{A}Q_{B}) ] \nonumber\\
&+& 4 f_{AB}' [\os^{\da\db} \bar{Q}_{A\da} \bar{Q}_{B\db} +
2i(\bar A\bar e_A \bar e_B- G_{5}{\bar\nu}_A{\bar\nu}_B)-2\sq i
{\bar F_1}\bar e_{A}\bar\nu_{B} \nonumber\\
&+& ( {\ovl W} Q_{A}Q_{B} +2{\ovl P} Q_{ A} L_{B}
+\sq {\ovl O} L_{A}L_{B})\nonumber\\
&-&2i t_{4}(\bar d_{A}\bar \nu_{B}+\bar u_{A} \bar e_{B})  +
  2i\sqrt{2} ( K\bar d_{A}\bar e_{B} - J_{1}\bar u_{A}\bar \nu_{B})
 ]  \label{MMFMS}\eea

We have suppressed $G_{321}$  indices and used a sub multiplet
naming convention specified in Section 2. and   and Table I  in
the Appendix.

In order that the  exchange of a Higgsino that couples to matter
with a given $B+L$ lead to a $B+L$ violating $d=5$ operator in the
effective theory at sub GUT
 energies it is necessary
that it have a nonzero contraction with  a conjugate (MSSM)
representation Higgsino that couples to a matter chiral bilinear
with a  $B+L$ different from the conjugate of the first $B+L$
value. Inspecting the above superpotentials one finds that only
$\{ \bar t_{(1)}, \bar t_{(2)}\}$ and $\{t_{(1)}, t_{(2)},
t_{(4)}\} $ satisfy this requirement. Terms containing the right
handed neutrinos $\bar \nu_A $ must
 be further processed to integrate out the heavy field $\bar \nu_A$
in favour of the the light neutrinos $\nu_A$ . This will introduce
an extra factor of $m^{\nu}_{Dirac}/M^{\bar\nu}_{Majorana}$ and
effectively lead to amplitudes suppressed like those of
 $d=6$ operators.
Thus  on integrating out the heavy triplet Higgs supermultiplets
one obtains the effective $d=4$ Superpotential for Baryon Number
violating processes   :

\bea  W_{eff}^{\Delta B\neq =0} = L_{ABCD} ({1\over 2}\epsilon Q_A
Q_B Q_C L_D) +R_{ABCD}  (\epsilon {\bar e}_A {\bar u}_B {\bar u}_C
{\bar d}_D) \eea
where the coefficients are

\bea L_{ABCD} =  {\cal S}_1^{~1}  h_{AB} h_{CD} + {\cal S}_1^{~2}
h_{AB} f_{CD} +
 {\cal S}_2^{~1}  f_{AB} h_{CD} + {\cal S}_2^{~2}  f_{AB} f_{CD} \eea

and \bea R_{ABCD} &=& {\cal S}_1^{~1}  h_{AB} h_{CD}
 - {\cal S}_1^{~2}  h_{AB} f_{CD} -
 {\cal S}_2^{~1}  f_{AB} h_{CD} + {\cal S}_2^{~2}  f_{AB} f_{CD} \nonumber \\
 &-& i{\sqrt 2} {\cal S}_1 ^{~4} f_{AB} h_{CD}
+i {\sqrt 2} {\cal S}_2 ^{~4} f_{AB} f_{CD} \eea

here ${\cal S}= {\cal T}^{-1} $ and ${\cal T} $ is the mass matrix
for $[3,1,\pm 2/3]$-sector  triplets :
 $W={\bar t} {\cal T} t
+...$, while

 \bea h_{AB} = 2 {\sqrt 2} h_{AB}' \qquad f_{AB} = 4
{\sqrt 2} f_{AB}' \eea

 This expression and the ``Clebsches" contained in it
, as well as the new baryon decay  channel  mediated by the
triplets ($t_{(4)}$)contained in
${\bf{\overline{\Sigma}_{126}(10,1,3)}}$   ( the same PS multiplet
that contains the Higgs field responsible for the right handed
neutrino Majorana mass), were given in \cite{alaps}.
 Previous work
\cite{babpatwil} on ${\bf{\overline{\Sigma}_{126}}}$ mediated
decay focussed  on the multiplets $t^{(2)},{\bar t}^{(2)} $ and
found  that there was no  contribution of $t^{(4)},{\bar t}^{(4)}$
in their models. This new   channel nominally strengthens the
emergent link between
 neutrino mass and baryon decay.
 Note however that $t_{(4)}$ couples only to the RR combinations
$ {\bar d} {\bar \nu} + \bar u \bar e $ and as such its exchange
 will contribute only to the RRRR channel which
, at least in SO(10), seems \cite{babpatwil}  generically
suppressed except at very large $tan \beta$.
 However the mixing in the
triplet mass matrix could also strengthen the effects of this
channel.

\subsection{Novel $d=5, \Delta(B+L)\neq 0$
Operators  via superheavy gaugino exchange ? }

A novel situation apparently arises in this GUT due to exchange of
superheavy gaugino Dirac multiplets that couple to matter
{\it{both}} via the gauge yukawa couplings of their gaugino part
{\it{and}} the superpotential couplings of their ($\oot$) chiral
components to the matter sector. Such gauginos are {\it{not}}
present in the case of $SU(5)$ As is evident from
eqn.(\ref{WSBFM}) the ${\bf{\oot}}$ submultiplet
 fields ${\bar E_2}[\bar 3,2,{-1/3}], {\bar{F_1}}[1,1,-2],J_1[3,1,4/3]$
  (which mix with the    superheavy
$SO(10)/SU(5) $ coset gauginos : see Section 2. and 3.) couple
only to terms containing at least one superheavy neutrino field
$\bar \nu_A$. Thus , to leading order in $M_U^{-1}$, the exchange
of such gaugino dirac multiplets will not lead to any $d=5$
operator with 4 light external fields.
However a puzzle remains.

 The superheavy neutrinos mix with the usual light neutrinos
  via   Dirac masses. So in
the effective theory one  trades  them for the light neutrinos by
using their equations of motion to leading order in their
(inverse) masses  ( effectively $\bar\nu
=-2(m_{\nu}^{Dirac}/M_{\bar \nu}^{Majorana}) \nu  + ...$).
 The chiral parts  ${\bar E_2}, {\bar F}_1 ,J_1$ of
 the Gauge   Dirac E,F,J  multiplets therefore  couple to
light neutrinos and another light matter field  with a  small
coupling  $\sim O(m^{Dirac}/M_{Majorana})$ .
 Exchange of the gaugino
dirac fermion between a gauge yukawa vertex and a $\oot\cdot
16\cdot 16 $ vertex can lead to effective operators involving 4
light matter fields of which at least one is a light neutrino and
one  is {\it{anti-chiral}}. This appears to violate
 the usual argument that in the
effective MSSM arising from a Susy GUT , supersymmetric  D terms
involving 4 light (mixed chiral and anti-chiral)
 fields must be $d\geq 6$ or equivalently that the $d=5, B,L$ violating terms
 are either of form $[QQQL]_F $ or $[\bar e\bar u\bar u\bar d]_F$.
Exchange  of $SO(10)/SU(5) $ coset gauginos peculiar to $SO(10) $
however appears to lead to (admittedly suppressed) $d=5$
chiral-anti-chiral operators with 4 light fields.
 These operators  arise once the Electroweak scale
vev that gives rise to neutrino Dirac masses is turned on. This
vev is {\it{smaller}} than $M_S$ and arises after soft
  susy breaking terms are included.  In this theory
$B-L$ is spontaneously broken giving rise to the Majorana mass for
  conjugate neutrinos (which was used to eliminate them in favour
of the SM neutrinos). Thus  perhaps the contradiction is not as
violent as it seems at first. We emphasize that there are no
analogous processes in SU(5) Susy GUTs since there the 12 coset
gauginos acquire partners from the purely AM type {\bf{24}} plets
which do not couple to the matter sector.

The couplings of the gauginos of SO(10) to the matter fields  are
easily computed  by adapting  the  PS reduction
 of the SO(10) covariant derivative
for the spinor $\bf 16$ \cite{alaps}:

\bea {\cal L}_{g-Y}&=& 2{i
g}{\bf{[}}{\tilde\psi}^{*}_{\kappa\alpha}
{\lambda}^{A}({t^{A}\over 2})_{\kappa}
^{~\mu}{\psi}_{\mu\alpha}+{{\widehat{\widetilde\psi}}}{}^{\mu
*}_{~~\dot\alpha} {\lambda}^{A} ((-{t_{A} \over 2})_{\mu\kappa})^*
\widehat\psi^{\kappa}_{~~\dot\alpha}
\nonumber\\
 &+&{{\widehat{\tilde\psi}}}{}^{\mu *}_{\dot\beta}
({{{\vec{\lambda}_{R}}\cdot \vec\sigma} \over 2})
_{\dot\beta}^{~~\dot\gamma}\widehat\psi_{\mu\dot\gamma}
+{\widetilde{\psi}_{\mu\beta}}^*({{\vec{\lambda}_{L} \cdot
\vec\sigma} \over 2})
_{\beta}^{~~\gamma}\psi_{\mu\gamma}{\bf{]}} \nonumber\\
&+&{{g  \over \sq }}
 [{ {\hat{\tilde\psi}^{\nu*}_{\dot\alpha}}}
{\lambda}^{\mu\nu\alpha}_{~~\dot\alpha}\psi_{\mu\alpha}+
\widetilde{\psi}_{\nu\alpha}^*
{\lambda}_{\mu\nu\alpha}^{~~\dot\alpha}
\widehat\psi^{\mu}_{~~\dot\alpha}]+ H.c \eea

The   terms carrying $B,L$ are  are \bea {\cal L}_{\Delta (B + L)
\neq 0} &=&
 \sqrt{2}g [   \widetilde{L}^{*}\bar J_{4}Q +
\widetilde{Q}^*J_{4}L ]
 -\sqrt{2}g[{\widetilde
{\bar{d}}}^{*}\bar{J}_{4}\bar{e}+{\widetilde
{\bar{u}}}^{*}\bar{J}_{4}\bar{\nu}+{\widetilde
{\bar{e}}}^{*}{J}_{4}\bar{d}+{\widetilde
{\bar{\nu}}}^{*}{J}_{4}\bar{u}]\\
 &+&{(g/\sq)}[{-\widetilde{\bar{d}}}^* {\ovl X_3}^{\alpha}L
-{\widetilde{\bar{u}}}^{*}\bar{E}_{(5)}L +{\widetilde
{\bar{e}}}^{*}{\ovl X_3}Q+{\widetilde {\bar{\nu}}}^{*}\bar{E}_{5}Q
+ \epsilon
 {\widetilde{\bar{d}}}^* E_5 Q + \epsilon
 {\widetilde{\bar{u}}}^* X_3 Q    ]\nonumber\\
&+& {(g/\sq)}[(\widetilde{L}^{*}(X_{3} \bar{d}-E_{5}\bar{u})-
\widetilde{Q}^{*}(X_3 \bar{e}-E_{5}\bar{\nu}) + \epsilon
\widetilde{Q}^{*} (\bar E_5 \bar{d}-{\ovl{X}}_{3}\bar{u}) ] +....
\nonumber\\
\eea

There are no $X[3,1,\pm 5/3]$ sector submultiplets in the $\oot$.
Thus we can  focus on just the $E[3,2,\pm{1/3}] $ and the
$J[3,1,\pm 4/3] $ sectors here. As discussed in Section 2.,
  the superheavy gauginos
${\ovl J_4}, E_5 $ mix with $\oot$ derived fermions $J_1[3,1,4/3]
$ and
 $ {\bar E}_2[\bar 3,2,-1/3] $.
  Examining eqn(\ref{MMFMS}) we see
that  $J_1,{\bar E_2}$  couple  only to operators involving at
 least one superheavy $\bar\nu$ field
($E_1\in \oot $  does not mix with E-gauginos):

\bea W_{FM}^{\Sigb}   =
 -8{\sqrt{2}}f_{AB}'
   [\bar E_{2}  Q_{B} + i J_{1}\bar u_{B}
 ]\bar\nu_{A}   =
[{\bar E}_2 Q_A + i J_1 {\bar u}_A] (f'M_{\bn}^{-1}
 m^{\nu D})_{AB} \nu_B +...\label{WFM126EJ}\eea

 Since $J_1$ couples to a $B=-1/3, L=1$
operator while ${\bar J}_4$  couples only to $B=1/3,L=-1$,   $J$
exchange does not lead to $B+L$ violation.  The E sector gaugino
i.e $E_5$ couples as

\be {g\over \sq} [ \epsilon {\tilde{\bar d}}^* E_5 Q + {\tilde
Q}^* E_5 \bn-{\tilde L}^* E_5{\bar u} ] \label{Egyuk}\ee

Only the first terms in (\ref{WFM126EJ},\ref{Egyuk}) are
 relevant and thus
we find the following effective lagrangian due to superheavy
gaugino exchange :

\be L_{\Delta(B+L)=2} = 4 g (f'(M^{\bar\nu})^{-1}
 m^{(\nu D)})_{AB}   {{\cal E}^{-1}}^{~2}_5
  [\epsilon {\widetilde{\bar d}}_C^* Q_C
   (Q_A\tilde\nu_B + {\tilde Q_A}\nu_B) ]
 \ee

where ${{\cal E}^{-1}}^2_5 $ is essentially the mass of the
exchanged gaugino times mixing factors written compactly interms
of the inverse of the relevant fermion mass matrix (in the
$E[3,2,\pm{1\over 3}]$ sector). By dressing this with MSSM
gauginos we obtain $\Delta B=\Delta L =1$ violating  4 fermi
vertices responsible for processes like

\be u_L d_L \longrightarrow  {\bar d}_L  +  {\ovl {\nu_L}} \ee

 This is  a  vertex quite
 distinct from the Higgsino mediated vertices since  it involves
 exchange of massive  gauginos between a chiral and
an anti-chiral vertex. It requires non zero external momenta for
the fermions and vanishes in the limit of zero external momenta.
Thus  the coefficient of the corresponding 4-fermi operators for B
violation in the effective lagrangian is $\sim M^{\nu
D}m_{Nucl}/M_X^2 M_S^2 $   where $M_S$ is the Susy breaking scale.
This magnitude seems hopelessly suppressed (relative even to gauge
boson exchange) to be observable. Nevertheless the contrast of its
structure with that of the standard $QQQL$ and $\bar u \bar u \bar
d \bar e$ operators perhaps warrants a more thorough investigation
of the conditions for the possibility of its appearance in the
effective theory.

\section{Fermion Mass Formulae}
 A vital issue for any SO(10) GUT is the type of predictions it
makes for the relations among the parameters of the (Type I and
Type II) seesaw mechanisms \cite{seesaw} by which  Neutrino masses
and mixings arise. From the coupling of  neutrinos to the
$\bf\oot$ we find that the Majorana
 mass matrix of the superheavy  neutrinos $\bar\nu_A$ is
( eqn.(\ref{MMFMS}))

 \be M^{\bn}_{AB}= -4i\sq
f'_{AB}<\Sigb^{(R+)}_{44}> =4\sq f'_{AB}\ssb \ee

 Similarly the Majorana mass matrix  for the left neutrinos
$\nu_A$ is ( eqn.(\ref{MMFMS})).

\be M^{\nu}_{AB}= 4\sq f'_{AB}<{\bar O}^{11}> =
 8i f'_{AB}<{\bar O}_{-}> \ee

where $<{\bar O}_{-}> $ is the small vev of the $SU(2)_L$ triplet
in the $(\ovt,3,1)_{\Sigb}$ induced by a tadpole that arises as a
consequence of $SU(2)_L$ breaking(see below).

In addition to this there is the Dirac mass which mixes the left
and right neutrinos  :

\be m^{\bn D}_{AB} = 2 \sq h'_{AB} <h^{(1)}_{ 2}> + 4 i {\sqrt 6}
f'_{AB} <h^{(2)}_{ 2}> \ee
 We must make the fine tuning $Det {\cal H} =0$
 necessary to keep a
pair of Higgs doublets $H_{(1)}, {\bar H}_{(1)}$ (which is to
develop the EW scale vev )  light. Then these doublets    will be
the left and right  null eigenstates of the mass matrix ${\cal
H}$. If the bi-unitary transformations responsible for
diagonalizing ${\cal H}^{\dagger} {\cal H} $ and ${\cal H}  {\cal
H}^{\dagger} $ are $U,{\overline U}$ i.e \be
Diag(m_H^{(1)},m_H^{(2)},....) =
 {\overline U}^T {\cal H} U
\nonumber \ee

then writing

\be h^{(i)} = U_{ij} H^{(j)} \qquad ;\qquad {\bar h}^{(i)} = {\bar
U}_{ij} {\bar H}^{(j)} \ee

where $H^{(j)},{\bar{H^{(j)}}}$ are the mass eigenstate doublets,
the contributions of any coupling in which $h^{(i)},{\bar
h}^{(i)}$ enter can be accounted for in the
 effective  MSSM below the heavy thresholds of the GUT
 just by replacing  $ h^{(i)} \rightarrow  \alpha_i H^{(1)},
{\bar h}^{(i)} \rightarrow  {\bar\alpha}_i {\bar H}^{(1)}$ where $
\alpha_i = U_{i1} , {\bar\alpha}_i =\bar U_{i1} $ and we have
numbered the massless doublet pair ``1". These components are
easily obtained from the normalized null eigenvectors $V,{\bar V}$
of ${\cal H}^{\dagger} {\cal H}$ and ${\cal H}{\cal H}^{\dagger}$
to be $ U_{i1}=V_i,{\bar U}_{i1}={\bar V}_i^*$. Thus the neutrino
Dirac mass matrix becomes

\be M^{\nu D}_{AB} = (2 \sq h'_{AB}\alpha_1 +
  4 i {\sqrt 6} f'_{AB} \alpha_2) v_u\ee

To obtain the final formula for the neutrino masses
 and mixings we must eliminate the $\bn $
fields which are superheavy and evaluate the tadpole that gives
rise to the Type II seesaw. The first step is standard.  As for
the $O(10,3,1)_{126}, {\overline O}({\overline{10}},3,1)_{\oot}$
vevs,  inspection of the mass spectrum (Table I in Appendix) and
eqns.(\ref{phSbH}-\ref{phSSb}) yields the relevant terms in the
superpotential as

\bea W_{FM}^{\Sigb} &=& M_O{\vec{\bar O}}\cdot  {\vec O}
-{{\bar\ga}\over {\sq}}H^{\al\da} \Phi_{44\da}^{\bet} {\bar
O}_{\al\bet} -{{\ga}\over {\sq}}H^{\al\da}
\Phi^{44\bet}_{\da} { O}_{\al\bet}\nonumber \\
&-&2\sq i \e ( \s_4^{~4\al\da} \Phi_{44\da}^{\bet}
 {\bar O}_{\al\bet}
+\Sigb_4^{~4\al\da} \Phi^{44\bet}_{\da}
 {O}_{\al\bet} )
\eea since $\s_4^{~4\al\dot 1} = {{-\sqt i}\over 2} {\bar
h}_{(3)}^{\al} ,{\overline{\s}}_4^{~4\al\dot 2} = {{\sqt i}\over
2} {  h}_{(2)}^{\al} , \Phi_{44\dot 1\al}= -\sq {\bar
h}^{(4)}_{\al}, \Phi^{44}_{\dot 2\al}= \sq {  h}^{(4)}_{\al}, $
one gets for the relevant terms  :

\bea
 M_O{\bar O}_{-} O_+  + {\bar O}_{-} {\sq}
({{{i\bar\ga} }} \bar\al_1 +2 i{\sqrt 3}\e \bar\al_3)
\bar\al_4 v_d^2 - { O}_{+} {\sq} ({{ i\ga} } \al_1 + i{2{\sqrt
3}}\e \al_2) \al_4 v_u^2 \eea

Thus the vev we need i.e $ <{\bar O}_{-}>$ is immediately
 determined to leading order in $M_W/M_U$ by by the equation for
$O_+$ as

\be <{\bar O}_{-}>  ={\sq}  ({{{i\ga} }} \al_1 + {2i\sqrt{3}}\e
\al_2) \al_4  ({{v_u^2}\over {M_O}}) \qquad \ee

 and $M_O$
can be read off from Table I to be $M_O= 2 (M + \eta (3a-p))$.

The quark and charged lepton mass matrices are

\bea
M^d &=&  (2\sq h'{\bar\al}_1 -
4 {\sqrt{2\over 3}}i f' {\bar\al}_2) v_d \\
M^u &=&  (2\sq h'\al_1 - 4 {\sqrt{2\over 3}}i
f' \al_2) v_u \\
M^l &=&  (2\sq h'{\bar\al}_1 +
 4 {\sqrt{6}}i f' {\bar\al}_2) v_d \\
\eea These formulae are now in a form ready to use for fitting the
fermion mass and mixing data  after lifting it via the RG
equations of the MSSM to the GUT scale.

\section{Discussion and Outlook}

In this paper we have calculated the complete superheavy spectrum
of the Minimal supersymmetric GUT along with the gauge and chiral
couplings of all MSSM multiplets in a readily accessible  form.
Partial calculations of these spectra and couplings
\cite{lee,alaps,bmsv,fm} have been published earlier but our
method is different from the computer based method
of\cite{lee,bmsv,fm} and is more complete, especially regarding
couplings. Being analytic and explicit it also allows us to trace
and resolve discrepancies arising within the computer based
approach. We used the calculated spectra to perform a preliminary
scan of the parameter space of the MSGUT as regards the magnitude
of the threshold corrections to two  crucial phenomenological
parameters of the MSGUT : the Weinberg angle at low energy and the
mass of the X lepto-quark gauge supermultiplet. We obtained a
result that is in sharp contrast with expectations in the
literature\cite{dixitsher} that precision RG calculations in
SO(10) are futile. On the contrary we find that the 1-loop GUT
threshold and gauge two loop contributions are modest but
significant. Thus, on the one hand, the basic GUT picture
suggested by the convergence of gauge couplings  in the MSSM is in
fact not destroyed by the contributions of the large number of
superheavy fields. On the other hand  extant precision
calculations that ignore threshold effects in SO(10) GUTs seem to
be of dubious validity. In particular the proper RG analysis of
the MSGUT taking into account EWRSB, all fermion masses and GUT
threshold effects still remains to be done. This calculation is
now being performed\cite{abmsv2}. In view of the other
phenomenological successes of renormalizable Susy SO(10)
\cite{babmoh,bsv,moh} and the unforseen correlations between
disparate phenomena like neutrino oscillations and nucleon decay
that have emerged\cite{babpatwil,alaps} the mildness and
calculability of threshold effects in the MSGUT is a most welcome
and promising development. Our preliminary scans of the MSGUT
parameter space (whose very feasibility - based on there being
just one ``sensitive" control parameter ($\xi$) - is a matter of
some astonishment) show that the threshold effects can potentially
narrow down the allowed regions of the MSGUT parameter space and
indicate correlations between the GUT scale and the $B-L$
violating scale which can be of crucial significance when cross
checking the particle physics phenomenology against cosmology.
 We have also argued that above the perturbative
  unification scale realistic renormalizable SO(10) GUTs are
  necessarily strongly coupled\cite{trmin,tas}. We have recently
  reported the results of 2-loop calculations of MSGUT
  RG equations {\it{above}} the SO(10) restoration scale which we
  used to show that the strong growth of the SO(10)
  coupling above $M_X$ cannot be evaded by taking shelter in a
  weakly coupled fixed point\cite{so10beta}. On the other hand our
  work\cite{trmin,tas} has shown that a scenario of
  a {\it{calculable}} dynamical
  symmetry breaking of  the GUT symmetry which utilizes the nearly
  exact supersymmetry at the GUT scale offers  rich possibilities
  for the significance of the new length scale
  associated with the
  condensation of $SO(10)/G_{321}$
  {\it{coset}} gauginos  implied by both holomorphic analysis and
  by the Konishi anomaly. The present calculations show the way
  for crossing the threshold and entering into the SO(10) regime
  in a controlled way. The emerging coherence of the low energy
  phenomenology, B and L violation,
   perturbative GUT structures (such as
  the natural R-parity preservation \cite{abs,abmrs},
 successful seesaw scenarios, leptogenisis etc) and exciting hints
 of deeper mysteries, perhaps unveilable\cite{trmin,tas},
 carry to our hopeful nostrils the  spoor
 of a grail perhaps  within reach.

\vspace{ .5 true cm} {\bf{Note Added}}
 \vspace{ .5 true cm}

After this paper was posted on the arXive as  hep-ph/0405074
the authors of hep-ph/0405300 claimed that the mass spectra listed
 in Appendix A were not internally consistent with the requirements of $SU(5)$ or
$SU(5)\times U(1)$ symmetry (at the special vevs  where $p=a=\pm \omega $).
 However this is   incorrect.
 The mass spectra we derived via a PS decomposition of SO(10)  organize straightforwardly
and  termwise  into appropriate $SU(5)$
 invariants  for SU(5) invariant  vevs as given in the Appendix B (added as above
to hep-ph/0205074v1).  This   term-wise  reorganization of several hundred $G_{123}$
 invariant mass terms into SU(5)  invariant mass terms is a   more stringent consistency
test than the   tests  of hep-ph/0405300 which are based on traces and determinants
   and valid only for their conventions.
 The phase conventions and field normalizations of hep-ph/0405300 are quite different
 from  our work. Thus the blind application of their trace  and determinant consistency
tests to our results  cannot but fail.
   We maintain unit  field normalizations throughout by using only unitary field
redefinitions of fields with canonical kinetic terms . Finally our
results for chiral spectra  also coincide, up to minor convention
related adjustments, with those obtained in a parallel computation
reported in \cite{bmsv}.

Finally we stress that our method\cite{alaps} yields {\it{all}}
coupling coefficients between both spinor and tensor irreps and
not just the tensor irrep ones relevant for masses and symmetry
breaking which were obtained using
 the method of \cite{heme}. Moreover we note that the most complex of the mass
matrices  given  here namely those of the Higgs doublets and
triplets   relevant
 for proton decay were already derived by us in  hep-ph/0204097v2(2003) \cite{alaps}

\vspace{ .5 true cm} {\bf{Further Note Added}}

 After version 2 of this paper (including the SU(5) reorganization given in Appendix B
above) was accepted for publication the authors of hep-ph/04050300
issued yet another preprint(hep-ph/0412348v1 ) this time claiming
that although our results pass  the SU(5) reorganization `test '
for $\sigma=\bar\sigma=0$ they   failed to do so for
 $\sigma=\bar\sigma\neq 0$ and that
  the counting of Goldstone modes and distinct mass
eigenvalues was, in their opinion incorrect. Further they claimed
that our results were inconsistent since they failed to pass
certain `trace and hermiticity tests ' that they had applied
successfully to their own results.   All these claims are
incorrect. Our results in fact pass all three tests. We have
issued a preprint showing this explcitly \cite{fukrebut}.
 Here we only remark that once the  super-Higgs effect for
$SO(10)\longrightarrow MSSM$ has been verified it is scarcely
feasible that the Goldstone-Higgs counting could fail for
$SO(10)\longrightarrow SU(5) $ since the latter is  a special case
of the same spectra ! However the reader can easily check that the
SU(5) singlet and 10plet mass matrices have zero determinant
confirming that the required Goldstone supermultiplets  ${\bf{1}}
+ {\bf{10}} +{\bf{\overline {10}}}$ are present. The demonstration
that the trace constraints and `hermiticity' tests of
hep-ph/0205300 are also satisfied is also straightforward once
proper account is taken of the difference in the phase conventions
of the two calculations. Details may be found in \cite{fukrebut}.
Finally as this paper goes to press the authors of
hep-ph/0412348v1 have reissued  the preprint hep-ph/0412348v2
in which all the  claims
 of the inconsistency of  our results are totally retracted.

 \vspace{ .5 true cm}
\section{Acknowledgments}
 \vspace{ .5 true cm}

It is a pleasure for C.S.A to acknowledge much correspondence  and
collaboration with B.~Bajc, A.~Melfo, G.~Senjanovi\'c and
F.~Vissani while calculating the Chiral spectra reported here.
C.S.A is also grateful to B. Bajc and Prof. D.R.T Jones for
correspondence related to RG analysis,
to Sumit Kumar for technical help with Appendix  B and
to the High Energy Group of the Abdus Salam ICTP, Trieste for hospitality while completing
Appendix B.  A.G. thanks H.R.I.,  Allahabad for its
facilities and hospitality. This work was done under Project
SP/S2/K-07/99 of  the Department  of Science and Technology of the Government of India.

 \vspace{ .5 true cm}
 {\bf {Appendix A : Tables
of masses and mixings }} \vskip .5 true cm
 \vspace{ .2 true cm}

In this appendix we collect our results for the chiral fermion/gaugino states,
masses and mixing matrices for the reader's convenience. Apart
from the discussion of gauge multiplet masses our results have
been obtained in parallel with and are compatible with those of
\cite{bmsv}, which, however , are computed with a different
normalization for the $\s,\Sigb$ fields resulting in a difference
between the mass and yukawa coupling parameters $M,\eta $ of these
multiplets in the two starting actions. Moreover certain minor
phase differences also exist between the definitions of
representative states  used by them and our definitions for the
same states (which follow directly from
 our consistent definitions of PS tensors from
  SO(10) submultiplets).
   Mixing matrix rows are labelled by barred
  irreps and columns by unbarred. Unmixed cases({\bf{i)}}) are given as Table
  I.

\begin{table}
$$
\begin{array}{l|l|l}
{\rm Field }[SU(3),SU(2),Y] &  PS  \qquad  Fields  & {\rm \qquad Mass}  \\
 \hline
 &&\\
 A[1,1,4],{} \bar A[1,1,-4] &{{\s^{44}_{(R+)}}\over
\sq}, {{\os_{44(R-)}}\over \sq}&
 2( M + \eta (p +3a +  6 \omega )) \\
 C_1[8,2,1],{} \bar C_1 [(8,2,- 1] &
\s_{\bn\alpha\dot1}^{~\bar\lambda},{}
 \Sigb_{\bn\alpha\dot 2}^{~\bar\lambda} &
 2 (-M + \eta (a+\omega)) \\
 &&\\
C_2[8,2,1],{}\bar C_2 [(8,2,- 1] &
\Sigb_{\bn\alpha\dot1}^{~\bar\lambda},{}
 \s_{\bn\alpha\dot 2}^{~\bar\lambda} &
 2 (-M + \eta (a-\omega)) \\
D_1[3,2,{7\over 3}],{}\bar D_1 [(\bar 3,2,- {7\over 3}] &
\Sigb_{\bn\alpha\dot1}^{~4},{}
 \s_{4\alpha\dot 2}^{~\bn} &
 2 (M + \eta (a+\omega)) \\
 &&\\
D_2[3,2,{7\over 3}],{}\bar D_2 [(\bar 3,2,- {7\over 3}]
 & \s_{\bn\alpha\dot1}^{~4},{}
 \Sigb_{4\alpha\dot 2}^{~\bn} &
 2 (M + \eta (a+3\omega)) \\
 &&\\
E_1[3,2,{1\over 3}],{}\bar E_1 [(\bar 3,2,- {1\over 3}] &
\Sigb_{\bn\alpha\dot 2}^{~4},{}
 \s_{4\alpha\dot 1}^{~\bn} &
 - 2 (M + \eta (a - \omega)) \\
K[3,1,-{8\over 3}],{}\bar K [(\bar 3, 1, {8\over 3}] & \Sigb_{\bn
4(R-)},{}
 \s^{\bn4}_{(R+)} &
 2 (M + \eta (a+p+ 2 \omega)) \\
 &&\\
L[6,1,{2\over 3}],{}\bar L [(\bar 6,1, -{2\over 3}] &
  (\Sigb_{\bm\bn}^{'(R0)},
 \s^{'\bm\bn}_{(R0)})_{\bm\leq\bn} &
 2 (M + \eta (p -a)) \\
& \Sigb'_{\bm\bn}=\Sigb_{\bm\bn},{} \bm\neq \bn&\\
&\Sigb'_{\bm\bm}={{\Sigb_{\bm\bm}}\over\sq}&\\
 &&\\
   M[6,1,{8\over 3}],{}{\ovl M} [(\bar 6,1, -{8\over 3}] &
  (\Sigb^{'(R+)}_{\bm\bn(R+)},{}
 \s^{'\bm\bn}_{(R-)})_{\bm\leq\bn} &
 2 (M + \eta (p -a + 2 \omega )) \\
N[6,1,-{4\over 3}],{}\bar N [(\bar 6,1, {4\over 3}] &
 (\Sigb_{\bm\bn}^{'(R-)},
  \s^{'\bm\bn}_{(R+)} )_{\bm\leq\bn} &
 2 (M + \eta (p -a-2\omega )) \\
 O[1,3,-2],{}\bar O [(1,3, +2] &
 {{{\vec\s}_{44(L)}}\over \sq},{}
 {{{{\vec{\Sigb}}^{44}_{(L)}}}\over \sq} &
 2 (M + \eta (3a-p)) \\
 &&\\
 P[3,3,-{2\over 3}],{}\bar P [\bar 3,3, {2\over 3}] &
 {\vec\s}_{\bm 4(L)},
  {\vec\Sigb}^{\bm 4}_{(L)}  &
 2 (M + \eta (a-p)) \\
  &&\\
W[6,3,{2\over 3}],{}{\overline W} [({\bar 6},3, -{2\over 3}] &
 {{{\vec\s'}_{\bm\bn(L)}}} ,
 {\vec\Sigb}^{\bm \bn}_{(L)}  &
 2 (M - \eta (a+p)) \\
I[3,1,{10\over 3}],{}\bar I [(\bar 3,1,- {10\over 3}] &
\phi_{~\bn(R+)}^4,{}
 \phi_{4(R-)}^{~\bn} &
 -2 (m + \lambda (p+a+4\omega)) \\
S[1,3,0] & \vec\phi^{(15)}_{(L)} & 2(m+\lambda(2a-p))\\
Q[8,3,0]& {\vec\phi}_{\bm(L)}^{~\bn}&
 2 (m - \lambda (a +p)) \\
U[3,3,{4\over 3}],{} \bar U[ \bar 3,3,-{4\over 3}] &
{\vec\phi}_{\bm(L)}^{~4},{} {\vec\phi}_{4(L)}^{~\bm}&
 -2 (m - \lambda (p-a)) \\
 &&\\
V[1,2,-3],{} \bar V[ 1,2,3] & {{{\phi}_{44\alpha\dot
2}}\over\sq},{} {{\phi^{44}_{\alpha\dot 1}}\over \sq}&
 2 (m  + 3 \lambda (a + \omega)) \\
B[6,2,{5\over 3}],{}\bar B [(\bar 6,2, -{5\over 3}] &
 (\phi_{\bm\bn\alpha\dot 1}',
 \phi^{'\bm\bn}_{\alpha\dot 2} )_{\bm\leq\bn} &
 -2 (m + \lambda (\omega -a )) \\
Y[6,2,-{1\over 3}],{}\bar Y [(\bar 6,2, {1\over 3}] &
 (\phi_{\bm\bn\alpha\dot 2}',
\phi^{'\bm\bn}_{\alpha\dot 1})_{\bm\leq\bn} &
 2 (m - \lambda (a+\omega )) \\Z[8,1,2],{} \bar Z[ 8,1,-2] & {\phi}_{~\bm(R+)}^{\bn}
{\phi}_{\bm(R-)}^{~\bn}&
 2 (m + \lambda (p-a)) \\
\end{array}
$$
\label{table I} \caption{{\bf{i)}}     Masses   of the unmixed
states in terms of the superheavy vevs . The $SU(2)_L$ contraction
 order is always $\bar F^{\alpha} F_{\alpha} $. The
 primed fields defined for $SU(3)_c$ sextets
  maintain unit norm. The absolute value
   of the expressions in the column ``Mass" is understood.}
\end{table}

\begin{table}
$$
\begin{array}{l|l|c|c}
{\rm Field }[SU(3),SU(2),Y] & \{S_3,S_2,S_1\}&S_W &S_X       \\
\hline A[1,1,4]  & \{0,0,{{12}/ 5}\}&9.6&12 \\
B[6,2,{5/ 3}] & \{5, 3,  5 \}&19.2&-6  \\
 C [8,2,1]  & \{6,4,{{12}/ 5}\}&4.8&-24\\
 D[3,2,{7/ 3}]  &\{1,{3/ 2},{{49}/ {10}}\}&10.8&21\\
E[3,2,{1/ 3}] & \{1,{3/ 2},{{ 1}/ {10}}\}&-8.4&-3  \\
F[1,1,2]& \{0, 0,{{ 3}/ {5}}\}&2.4&3  \\
G[1,1,0]& \{0, 0, 0\}&0&0  \\
h[1,2,1]& \{0, {1/ 2},{{ 3}/ {10}}\}&-3.6&3  \\
I[3,1,{10/ 3}]& \{  {1/ 2},0, 5 \}&22.8& 21   \\
J[3,1,{4/ 3}]&\{  {1/ 2},0,{{ 4}/ { 5}}\}&6&0 \\
K[3,1, {8/ 3}]&\{{1/ 2},0,{{  16}/ { 5}}\}&15.6&12\\
L[6,1,{2/ 3}] &\{{5/ 2},0,{{ 2}/ {5}}\}&15.6&-18\\
M[6,1,{8/ 3}]  &\{{5/ 2},0,{{ 32}/ {5}}\}&39.6&12\\
N[6,1, {4/ 3}]  &\{{5/ 2},0,{{ 8}/ {5}}\}&20.4&-12\\
O[1,3, 2]  &\{ 0,2,{{ 9}/ {5}}\}&-12&15 \\
P[3,3, {2/ 3}] &\{{3/ 2}, 6,{{  3}/ {5}}\}&-46.8&9\\
Q[8,3,0]& \{ 9,16,0 \}&-103.2&-24  \\
R[8,1, 0] & \{3, 0,  0 \}&16.8&-24\\
S[1,3,0] & \{ 0,2, 0 \}&-19.2&6 \\
t[3,1,{2/ 3}]&\{{1/ 2},0,{{1 }/ {5}}\}&3.6&-3\\
U[3,3,{4/ 3}]  &  \{{3/ 2},6,{{12}/ {5}}\}&-39.6&18 \\
V[1,2,3] & \{0,{ 1/ 2}, {{  27}/ { 10}}\}&6&15  \\
W[6,3,{2/ 3}]  & \{{15/ 2},12 ,{{  6}/ { 5}}\}&-68.4&-18\\
X[3,2,{5/ 3}]&\{1,{3/ 2}, {{5 }/ {2}}\}&1.2&9\\
Y[6,2, {1/ 3}]  & \{ 5, 3,{{1 }/ 5}\}&0&-30\\
Z[8,1,2] &\{ 3, 0, 24/ 5 \}&36&0   \\
\end{array}
$$
\label{table II} \caption{ Index values for the 26 different
chiral multiplet types (used in the threshold corrections). Except
for Q,R,S all other reps come in complex pairs.$ S_W = 4S_1-9.6 S_2
+5.6 S_3, S_X= 5 S_1+3 S_2-8 S_3 $ are the combinations that enter
the threshold corrections to $Sin^2\theta_W$ and to $Log_{10} M_X$
}
\end{table}

 {\bf{ ii)\hspace{ 1.0 cm} Chiral Mixed states}}\hfil\break

\vspace{ .3 cm}

 {\bf{a)}}$ [8,1,0](R_1,R_2)\equiv (\hat\phi_{\bm}^{~\bn},\hat\phi_{\bm
(R0)}^{~\bn})  $

 \be
{\cal{R}} = 2 \left({\begin{array}{cc} (m-\lambda a ) &
-\sqrt{2}\lambda\omega \\ -\sqrt{2}\lambda\omega & m+\lambda( p-a)
\end{array}}\right)
\ee

\be m_{R_{\pm}}=   |{\cal R}_{\pm}| = |2 m [1 +({\pt \over 2} -
\at) \pm \sqrt{({\pt\over 2})^2 + 2 \omt^2}]| \ee
 The  corresponding
eigenvectors can be found by diagonalizing the matrix ${\cal
R}{\cal R}^{\dagger}$. \vspace{ .3 cm}

{\bf{b)}}\hspace{ 1.0cm}  $ [1,2,-1]({\bar h}_1,\bh_2,\bh_3,\bh_4)
\oplus     [1,2,1](h_1,h_2,h_3,h_4) $\hfil\break
 $.\qquad\qquad\equiv
(H^{\alpha}_{\dot 2},\Sigb^{(15)\alpha}_{\dot 2},
\s^{(15)\alpha}_{\dot2},{{\phi_{44}^{\dot 2\alpha}} \over
\sq})\oplus (H_{\alpha {\dot 1}},\Sigb^{(15)}_{\alpha \dot1},
\s^{(15)}_{\alpha\dot 1}, {{\phi^{44\dot 1}_{\alpha}} \over \sq})
 $

\bea {\cal{H}}=\left({\begin{array}{cccc} -M_{H} &
+\overline{\gamma}\sqrt{3}(\omega-a) & -{\gamma}\sqrt{3}(\omega+a)
& -{\bar{\gamma}{\bar{\sigma}} }\\
 -\overline{\gamma}\sqrt{3}(\omega+a) & 0 & -(2M+4\eta(a+\omega)) &
0\\
\gamma\sqrt{3}(\omega-a) & -(2M+4\eta(a-\omega)) & 0 &
-{2\eta\overline{\sigma}\sqrt{3}}\\
  -{\sigma\gamma } &
-{2\eta\sigma\sqrt{3}} & 0 & {-{2m}+6\lambda(\omega-a)}
\end{array}}\right) \nonumber \eea

The above matrix is to be diagonalized after imposing the fine
tuning condition $Det {\cal H} =0$ to keep one pair of doublets
light.

\vspace{ .5 true cm}

 {\bf{c)}} $[\bar 3,1,{2\over 3}]
(\bt_1,\bt_2,\bt_3,\bt_4,\bt_5) \oplus [3,1,-{2\over 3}]
(t_1,t_2,t_3,t_4,t_5)$\hfil\break $.\qquad\qquad\equiv (H^{\bm
4},\Sigb_{(a)}^{\bm 4}, \s^{\bm 4
}_{(a)},\s^{\bm4}_{R0},\phi_{4(R+)}^{~\bm}) \oplus (H_{\bm
4},\Sigb_{(a)\bm 4}, \s_{\bm 4 (a)},\Sigb_{\bm
4(R0)},\phi_{\bm(R-)}^{~4}) $

\bea {\cal{T}}= \left({\begin{array}{ccccc} M_{H} &
\overline{\gamma}(a+p) & {\gamma}(p-a) & {2\sqrt{2}i
\omega{\bar\gamma}} & i\bar{\sigma}\bar{\gamma}\\ \bar\gamma(p-a)
& 0 & 2M & 0 & 0\\ \gamma(p+a) & 2M & 0 & 4\sqrt{2}i \omega\eta &
2i{\eta\overline{\sigma}}\\ -2\sqrt{2}i \omega\gamma  &
-4\sqrt{2}i \omega\eta & 0 & 2M+2\eta{p}+2\eta a &
-2\sqrt{2}{\eta\overline{\sigma}}\\ i\sigma\gamma & 2i \eta\sigma
& 0 & 2\sqrt{2}\eta\sigma & -2m - 2\lambda(a+p-4 \omega)
\end{array}}\right)
\nonumber\eea

\vspace{ .5 cm}

 {\bf{iii)  Mixed gauge chiral .}}

{\bf{a)}}$ [1,1,0] (G_1,G_2,G_3,G_4,G_5,G_6) \equiv
(\phi,\phi^{(15)},\phi^{(15)}_{(R0)},{{\s^{44}_{(R-)}}\over \sq},
{{\Sigb_{44((R+)}}\over \sq}, {{{\sq \lambda^{(R0)} -
{\sqrt{3}}\lambda^{(15)}}\over {\sqrt{5}}}})$

\bea {\cal{G}}= 2\left({\begin{array}{cccccc} m&0 &
\sqs\la\om & {{i\e\ssb}\over \sq}&{-i\e\sss\over \sq}&0\\
0& m + 2 \la a & 2\sq\la\om& i\e\ssb\sqtt &-i\e\sss\sqtt&0\\
\sqs\la\om&2\sq\la\om&m+\la(p+2a)& -i\e\sqt\ssb & i\sqt\e\sss&0\\
{{i\e\ssb}\over\sq}& i\e\ssb\sqtt&-i\e\sqt\ssb&0&
M+\e(p+3a -6\om)&{{\sqf g\sss^*}\over  2 }\\
{{-i\e\sss}\over\sq}& -i\e\sss\sqtt&i\e\sqt\sss&
M+\e(p+3a -6\om)&0&{{\sqf g \ssb^*}\over 2}\\
0&0&0&{{\sqf g\sss^*}\over 2}&{{\sqf g\ssb^*}\over 2}&0
\end{array}}\right)
\nonumber\eea

{\bf{b)}}  $[\bar 3,2,-{1\over 3}](\bar E_2,\bar E_3,\bar E_4,\bar
E_5) \oplus [3,2,{1\over 3}](E_2,E_3,E_4,E_5)$\hfil\break
$.\qquad\qquad \equiv (\Sigb_{4\alpha\dot 1}^{\bm}, \phi^{\bm
4}_{(s)\alpha\dot 2} , \phi^{(a) \bm 4}_{\alpha\dot
2},\lambda^{\bm 4}_{\alpha\dot 2}) \oplus  (\s_{\bm\alpha\dot
2}^4,\phi_{\bm 4\alpha\dot 1}^{(s)}, \phi_{\bm 4\alpha\dot
1}^{(a)},\lambda_{\bm\alpha\dot 1}) $

\bea {\cal{E}}= \left({\begin{array}{cccc} -2(M+\e(a-3\om))&
-2\sq i\e\sss&2i\e\sss&ig\sq\ssb^*\\
2i\sq\e\ssb&-2(m+\la(a-\om))&-2\sq\la\om&2g(a^*-\om^*)\\
-2i\e\ssb&-2\sq\la\om&-2(m-\la\om)&\sq g(\om^*-p^*)\\
-ig\sq\sss^*&2g(a^*-\om^*)&g\sq(\om^*-p^*)&0
 \end{array}}\right)
\eea

{\bf{c)}}$ [1,1,-2](\bar F_1,\bar F_2, \bar F_3)
 \oplus [1,1,2](F_1,F_2,F_3) $ \hfil\break
$.\qquad\qquad \equiv
(\Sigb_{44(R0)},\phi^{(15)}_{(R-)},\lambda_{(R-)})
  \oplus (\s^{44}_{(R0)},\phi^{(15)}_{(R+)},\lambda_{(R+)}) $ .

\bea {\cal{F}}= \left({\begin{array}{ccc} 2(M+\e(p+3 a))&
-2i\sqt\e\sss&-g\sq\ssb^*\\
2i\sqt\e\ssb&2(m+\la(p+2a))& {\sqrt{24}}ig\om^*)\\
 -g\sq\sss^*&-{\sqrt{24}}ig\om^*&0
 \end{array}}\right)
\eea

{\bf{d)}} $[\bar 3,1,-{4\over 3}](\bar J_1,\bar J_2,\bar J_3,\bar
J_4) \oplus [3,1,{4\over 3}](J_1,J_2,J_3,J_4)$ \hfil\break
$.\qquad\qquad\equiv (\s^{\bm4}_{(R-)},\phi_4^{\bm},
\phi_4^{~\bm(R0)},\lambda_4^{~\bm}) \oplus
(\Sigb_{\bm4(R+)},\phi_{~\bm}^4,
\phi_{\bm(R0)}^{~4},\lambda_{\bm}^4)$

\bea {\cal{J}}= \left({\begin{array}{cccc} 2(M+\e(a+p-2\om))&
-2\e\ssb&2\sq \e\ssb&-ig\sq\sss^*\\
2\e\sss&-2(m+\la a)&-2\sq\la\om&-2ig\sq a^*\\
-2\sq\e\sss&-2\sq\la\om&-2(m+\la(a+p))&-4i g\om^*\\
-ig\sq\ssb^*&2\sq ig a^*&4i g\om^*&0
 \end{array}}\right)
\eea

{\bf{e)}}$ [3,2,{5\over 3}](\bar X_1,\bar X_2,\bar X_3) \oplus
[3,2,-{5\over 3}](X_1,X_2,X_3)\hfil\break .\qquad\qquad \equiv
(\phi^{(s)\bm4}_{\alpha\dot 1} , \phi^{(a)\bm4}_{\alpha\dot 1}
,\lambda^{\bm4}_{\alpha\dot 1}) \oplus(\phi_{\bm4\alpha\dot
2}^{(s)}, \phi_{\bm4\alpha\dot 2}^{(a)},\lambda_{\bm4\alpha\dot
2})
  $

\bea {\cal{X}}= \left({\begin{array}{ccc} 2(m+\la(a+\om))&
-2\sq \la \om &-2g(a^*+\om^*)\\
-2\sq \la \om &2(m+\la \om)& {\sq}g(\om^* +p^*)\\
 -2 g(a^* +\om^*) &\sq g(\om^* + p^*)&0
 \end{array}}\right)
\eea

\vspace{ .5 true cm}
 {\bf {Appendix B }}:  ${\bf{SU(5)}}\times {\bf{U(1)}}$  {\bf{Reassembly Crosscheck}} \vskip .5 true cm
 \vspace{ .2 true cm}

 Given the  complexity of the spectra and couplings derived here  it would be useful to  have a method of cross checking the internal consistency of our results. A stringent  check is provided by verifying   that at special values of the vevs  i.e
\be
 p=a=\pm \omega \qquad
\ee
where the unbroken symmetry includes SU(5) the MSSM labelled mass spectra and couplings given in Appendix A  do  indeed reassemble  into  SU(5) invariant form.   For the mass spectra this is fairly  straightforward  to  check and is reported explicitly below.  A similar calculation\cite{aulsumit} for the super potential couplings is much more tedious but furnishes  an $SO(10)- SU(5) \times U(1) $ analog of the ``SO(10)-PS Clebsches '' reported here .

The decomposition of the chiral multiplets of the MSGUT into  $SU(5)\times U(1)$ multiplets and of those into MSSM multiplets (named as per the alphabetic convention of Appendix A) is   given below. The only complication is that certain MSSM multiplet types  occur in several copies and (orthogonal) mixtures of these are present in the different SU(5) mutiplets. Thus, for instance, the 210 contains a 24 and a 75 of SU(5)  both of which contain  mixtures of the $G_{123}$ multiplets $R_1(8,1,0)$ and $R_2(8,1,0)$. These mixtures must be orthogonal and must be precisely the eigenstates of the mass matrices in this $G_{123}$ sector which have the same   masses as the rest of the $G_{123}$ submultiplet sets  within the  24-plet and 75-plets as two wholes. The fact that this follows in every case from our results   appears to confirm their reliability.  The decompositions we need are :

\begin{eqnarray}
{\bf{H}} &=& 10 = 5_1 + {\bar{5}}_{-1}\nonumber \\
 {\bf{\Sigma}} &=& 126 = 1_{-5} (G_4) + \bar 5_{-1} + 10_{-3} + \overline{15}_3
+ 45_1 + \overline{50}_{-1}  \nonumber \\
  \bar 5_{-1} &=& \bar h_3 (1,2,-1) + \bar t_{3,4} (\bar
3,1,\frac{2}{3})  \nonumber \\
 10_{-3} &=& F_1 (1,1,2) + \bar J_1 (\bar 3,1,-\frac{4}{3}) + E_2
(3,2,\frac{1}{3}) \nonumber \\
 \overline{15}_3 &=& O(1,3,-2) + \bar E_1 (\bar 3,2,-\frac{1}{3}) +
\bar N (\bar 6,1,\frac{4}{3})  \nonumber \\
 45_1 &=& h_3 (1,2,1) + t_3 (3,1,-\frac{2}{3}) +
P(3,3,-\frac{2}{3}) + \bar K (\bar 3,1,\frac{8}{3}) + \bar D_1
(\bar 3,2,-\frac{7}{3})\nonumber \\
&+& \bar L (\bar 6,1,-\frac{2}{3}) + C_1
(8,2,1)  \nonumber \\
 \overline{50}_{-1} &=& A(1,1,4) + \bar t_{3,4}(\bar 3,1,\frac{2}{3})
+ D_2 (3,2,\frac{7}{3}) + W(6,3,\frac{2}{3}) + {\overline M} (\bar
6,1,-\frac{8}{3}) + \overline C_2 (8,2,-1)  \nonumber \\
{\bf{\Sigb}} &=& \oot = 1_{5} (G_5) +   5_{1} + {\overline{10}}_{3} +  {15}_{-3}
+ {\overline{45}}_{-1} +  {50}_{1}  \nonumber \\
    5_{ 1} &=&   h_2 (1,2,1) +   t_{2,4} (  3,1,-\frac{2}{3})  \nonumber \\
 {\overline{10}}_{3} &=& {\bar{F}}_1 (1,1,-2) +  J_1 (  3,1,\frac{4}{3}) +
 {\bar E}_2(\bar 3,2,-\frac{1}{3}) \nonumber \\
  {15}_{-3} &=& {\bar O}(1,3,2) +  E_1 ( 3,2,\frac{1}{3}) +
 N ( 6,1,-\frac{4}{3})  \nonumber \\
 {\overline{45}}_{-1} &=& \bar h_2 (1,2,-1) + \bar t_2 (3,1,\frac{2}{3}) +
{\bar P}(\bar 3,3,\frac{2}{3}) +   K (  3,1,-\frac{8}{3}) +   D_1
( 3,2,\frac{ 7}{3})\nonumber \\
&+&   L (  6,1,\frac{ 2}{3}) + \bar C_1
(8,2,-1)  \nonumber \\
  {50}_{ 1} &=& \bar A(1,1,-4) +  t_{2,4}(  3,1,-\frac{2}{3})
+ \bar D_2 (\bar 3,2,-\frac{7}{3}) +\overline W(\bar 6,3,-\frac{2}{3}) + {  M} (
6,1,\frac{ 8}{3}) +   C_2 (8,2, 1)  \nonumber \\
 {\bf{\Phi}} &=& 210 = 1_0 + 5_{-4} + \bar 5_4 + 10_2 + \overline{10}_{-2} +
24_0 + 40_2 + \overline{40}_{-2} + 75_0\nonumber \\
 1_0 &=& G_{1,2,3}  \nonumber \\
   5_{-4} &=& h_4 (1,2,1) + t_5 (3,1,-\frac{ 2}{3})\nonumber \\
  \bar 5_{4} &=& \bar h_4 (1,2,-1) + \bar t_5 (\bar 3,1,\frac{ 2}{3})\nonumber \\
 10_2 &=& F_2 (1,1,2) + \bar J_{2,3} (\bar 3,1,-\frac{4}{3}) +
E_{3,4} (3,2,\frac{1}{3})  \nonumber \\
 \overline{10}_{-2} &=& \bar F_2 (1,1,-2) + J_{2,3} (3,1,\frac{4}{3}) +
\bar E_{3,4} (\bar 3,2,-\frac{1}{3}) \nonumber \\
 24_0 &=& (1,1,0) G_{1,2,3} + S (1,3,0) + X_{1,2}
(3,2,-\frac{ 5}{3}) + \bar X_{1,2} (\bar 3,2,\frac{5}{3}) + R_{1,2}
(8,1,0) \nonumber \\
 40_2 &=& V (1,2,-3) + E_{3,4} (3,2,\frac{1}{3}) + \bar J_{2,3}
(\bar 3,1,-\frac{4}{3}) + \bar U (\bar 3,3,-\frac{ 4}{3}) + Z
(8,1,2) + \bar Y (\bar 6,2,\frac{1}{3})  \nonumber \\
\overline{40}_{-2} &=& \bar V (1,2,3) + \bar E_{3,4} (3,2,-\frac{1}{3}) +
 J_{2,3}( 3,1,\frac{4}{3}) +  U (3,3,\frac{4}{3}) + \bar Z
(8,1,-2) + Y (6,2,-\frac{1}{3})  \nonumber \\
 75 &=& (1,1,0) G_{1,2,3} + I (3,1,\frac{10}{3}) +  \bar I (\bar 3,1,-\frac{10}{3}) + X_{1,2}
(3,2,-\frac{5}{3}) +\bar
X_{1,2} (\bar 3,2,\frac{5}{3}) \nonumber \\
  & + &   B(6,2,\frac{5}{3}) + \bar B (\bar
6,2,-\frac{5}{3}) + R_{1,2} (8,1,0) + Q
(8,3,0)
\end{eqnarray}

If we insert $a=-\omega=p $ in the mass matrices of Appendix A we
find that, after diagonalizing the mass matrices of the
submultiplets that mix, the resultant spectra   group   precisely
as indicated by the decompositions above with all the subreps of a
given SU(5) irrep obtaining the same mass. One obtains the $ SU(5)
$ invariant mass terms :

\begin{eqnarray}
  && 2(M  +  10\eta p) 1_{\Sigma} 1_{\overline\Sigma} + 2(M + 4 \eta p)
  \bar 5_{\Sigma}  5_{\overline\Sigma}  + 2(M - 2\eta
p) \overline{50}_{\Sigma} 50_{\overline\Sigma} \nonumber \\
&+& 2(M + 4\eta p) 10_{\Sigma} \overline{10}_{\overline\Sigma} + 2(M + 2\eta p)
\overline{15}_{\Sigma}  15_{\overline\Sigma}
+ 2M 45_{\Sigma} \overline{45}_{\overline\Sigma}  \nonumber \\
 & + & M_H  \bar 5_H  5_H + ( m + 6\lambda p) (1_{\Phi})^2 +
 2(m + 6 \lambda p) 5_{\Phi} \bar 5_{\Phi} + 2(m
  + 3  \lambda p) 10_{\Phi} \overline{10}_{\Phi} \nonumber \\
 &+& (m + \lambda p)
 (24_{\Phi})^2 + 2m 40_{\Phi} \overline{40}_{\Phi} + (m - 2\lambda p)
 (75_{\Phi})^2 \nonumber \\
  &+& 2\eta\sqrt 3 (\overline\sigma (\bar 5_{\Sigma}
 5_{\Phi} + 10_{\Sigma} \overline{10}_{\Phi})
  +  \sigma ( 5_{\overline\Sigma} \bar 5_{\Phi} + \overline{10}_{\overline\Sigma}
 10_{\Phi}))   \\  &+&
 2\sqrt 3 p(\gamma \bar 5_{\Sigma} 5_H +
 \bar{\gamma} \bar 5_H 5_{\overline \Sigma})
+ 2i {\eta\sqrt 5} ( \sigma  1_{\overline\Sigma} 1_{\Phi}
 - \overline\sigma  {1}_{\Sigma}1_{\Phi})
+ {\gamma }   \sigma  {\bar 5}_{\Phi} 5_{H}
 +{\overline\gamma} \overline\sigma  {\bar 5}_{H}5_{\Phi}
 \nonumber\label{su5reass}
\end{eqnarray}

Where every $SU(5)$ invariant has been normalized so that the individual
$G_{123}$ sub-rep masses can be read off directly from the coefficient of the
invariant  for complex SU(5) representations which pair into Dirac supermultiplets and
is 2 times the coefficient for the real representations which remain unpaired
Majorana/Chiral supermultiplets. For $a=-\omega=p , \sigma=\bar\sigma=0$ the 20
Goldstone supermultiplets $G,J,\bar J, F,\bar F, E,\bar E)$ of the coset
 $SO(10)/SU(5)\times U(1) $ remain heavy (see the gauge-chiral super-Higgs
mass formulae in Section 2) as they should since they are eaten in
 the spontaneous breaking $SO(10) \longrightarrow
SU(5)\times U(1) $ while the 12 fields in the $\{X_3,\bar X_3\}$ multiplets lose
 their mass terms with $\{X_{1,2},\bar X_{1,2}\}$ since they form part of the
unbroken SU(5) gauge supermultiplet. When $a=\omega=p $ i.e for
flipped SU(5),
  the roles of the $\{X_3,\bar X_3\} $ and $\{E_5,\bar E_5\}$ gauge multiplets
are interchanged, with the E's remaining massless and the X's becoming heavy,
so that one obtains the SU(5) invariant groupings corresponding to
the ``flipped'' $SU(5)\times U(1) \subset SO(10) $  embedding.
Note that this successful SU(5) reassembly is a much more fine-grained
consistency test than any overall trace or determinant test.

\end{document}